\documentclass[fleqn,usenatbib]{mnras}

\usepackage{newtxtext,newtxmath}

\usepackage[T1]{fontenc}

\DeclareRobustCommand{\VAN}[3]{#2}
\let\VANthebibliography\thebibliography
\def\thebibliography{\DeclareRobustCommand{\VAN}[3]{##3}\VANthebibliography}

\usepackage{soul}
\usepackage{graphicx}	
\usepackage{amsmath}	

\title[A gap in the double white dwarf separation distribution]{
A gap in the double white dwarf separation distribution caused by the common-envelope evolution: astrometric evidence from {\it Gaia}}

\author[Korol et al]{Valeriya Korol$^{1}$\thanks{These authors contributed equally.}\thanks{E-mail: \href{mailto:korol@star.sr.bham.ac.uk}{korol@star.sr.bham.ac.uk} (VK)}, Vasily Belokurov$^{2,3}$ \footnotemark[1]\thanks{E-mail: \href{mailto:vasily@ast.cam.ac.uk}{vasily@ast.cam.ac.uk} (VB)}, Silvia Toonen$^{1,4}$\footnotemark[1]\thanks{E-mail: \href{mailto:Toonen@uva.nl}{toonen@uva.nl} (ST)}, \\
$^1$Institute for Gravitational Wave Astronomy \& School of Physics and Astronomy, University of Birmingham, Birmingham, B15 2TT, UK\\
$^2$Institute of Astronomy, Madingley Rd, Cambridge, CB3 0HA \\
$^3$Center for Computational Astrophysics, Flatiron Institute, 162 5th Avenue, New York, NY 10010, USA\\
$^3$Anton Pannekoek Institute for Astronomy, University of Amsterdam, 1090 GE Amsterdam, The Netherlands
}

\begin{document}

\defcitealias{GF21}{GF21}
\defcitealias{B20}{B20}


\maketitle

\label{firstpage}

\begin{abstract}
The trajectory of the center of light of an unresolved binary is different from that of its center of mass. Binary-induced stellar centroid wobbling can therefore be detected as an excess in the goodness-of-fit of the single-star astrometric model. We use reduced $\chi^2$ of the astrometric fit in the {\it Gaia} Early Data Release 3 to detect the likely unresolved double white dwarfs (DWDs). Using parallax-based distances we convert the excess of reduced $\chi^2$ into the amplitude of the centroid wobble $\delta a$, which is proportional to the binary separation $a$. The measured $\delta a$ distribution drops towards larger wobble amplitudes and shows a break around $\delta a \approx 0.2$ where it steepens. The integral of the distribution yields DWD fraction of $6.5 \pm 3.7$ per cent in the range $0.01 < a\,(\text{au}) < 2$. Using synthetic models of the Galactic DWDs we demonstrate that the break in the $\delta a$ distribution corresponds to one side of a deep gap in the DWD separation distribution at around $a\approx 1$\,au. Model DWDs with separations less than several au shrink dramatically due to (al least one) common envelope phase, reshaping the original separation distribution, clearing a gap and creating a pile-up of systems with $a\approx 0.01$\,au and $\delta a < 0.01$. Our models reproduce the overall shape of the observed $\delta a$ distribution and its normalisation, however the predicted drop in the numbers of DWDs beyond the break is steeper than in the data.
\end{abstract}


\section{Introduction}

At birth, binary star separations are typically large, with the period distribution of doubles on the main sequence (MS) peaking at $10^5$ days \citep[see e.g.][]{Raghavan2010,Badenes2018}. Thus, to match observed compact orbital configurations of systems with evolved companions, such as e.g. cataclysmic variables, an evolutionary process is required that would efficiently remove energy and angular momentum from the binary. Linked to the stability of the mass-transfer as one of the stars evolves from the MS and fills its Roche lobe, common envelope (CE) evolution is the hypothesized phase introduced to facilitate the necessary orbit shrinkage \citep[][]{Ostriker1973,Paczynski1976,Huel1976}. Sorting astrophysical phenomena by the decreasing ratio of their importance to the availability of direct observational constraints, CE ought to be close to the very top. It is for example invoked in both single-degenerate and double-degenerate supernovae type Ia explosion scenarios \citep[][]{Whelan1973,Iben1984,Webbink1984}. Formation channels for the gravitational wave sources that are pairs of compact stellar remnants rely on it too \citep[e.g.][]{Belczynski2002,Abadie2010,Nelemans01a}.

The enormous efforts of the community to understand the multi-spatial multi-timescale CE-phase in detail is therefore no surprise \citep[for a review][]{Ivanova2013}. Hydrodynamical simulations have investigated the physics of the CE-phase and 
have shed many insights on the phenomenon,
for example regarding the role of recombination energy \citep{Nan15,Rei20, San20}, 
the magnetic field \citep{Reg95, Nor07, Ohl16}, 
accretion \citep{Mac15,Cha18}, 
dust \citep{Gla18, Iac20}, 
jets \citep{Shi19,Lop21},
or the initial conditions \citep{Pas12,Iva16,Iac18, Mac18,Rei19,Gla21}. 
However, for a long time, the simulations failed to unbind the entire envelope. Successful CE evolution has now been achieved under certain conditions; for high mass systems \citep{Law20} or due to inclusion of recombination energy  \citep{Nan15,Rei20,San20}. However with the current lack of predictive power from these methods, evolutionary modelling of binary star systems relies on simplistic models for CE-evolution, that leave much to be desired. The models typically consider the energy budget  \citep{Paczynski1976, Webbink1984, Liv88} or the angular momentum budget \citep{Nelemans2000}. Typically the uncertainty in the outcome of the common-envelope phase is the most uncertain factor in the formation of compact binaries.

 On the observational side, specifically for the low- and intermediate-mass progenitors, constraints on CE-evolution come from the measurements of orbital separations of white dwarf - main-sequence pairs \citep{Zorotovic2010, Toonen13, Camacho2014} and double white dwarf (DWD) systems \citep{Nelemans01a, Nelemans2005, van06}.  
In particular compact DWDs are extremely scarce and incomplete \citep[see e.g.][]{Maxted1999,Badenes2012,Toonen2017,Maoz2018}.

In the past, two observational techniques have been used successfully to discover close DWDs. Most of the systems known to date, either with regular WD companions, with masses similar to observed single WDs, or with extremely low-mass (ELM) WDs ($M<0.3\,$M$_\odot$), which cannot be formed through single-star evolution within a Hubble time, have been discovered through radial velocity variations \citep[see e.g.][]{Marsh1995,Napiwotzki2001,Brown2010,Breedt2017,Napiwotzki2020, Brown2020}. Selection effects in the spectroscopically identified DWD samples come from two sources: that of the input catalogue and that of the spectroscopic follow-up program. While the survey selection effects can be pinned down, the input catalogue for the non-ELM surveys (such as  the Supernova Ia Progenitor surveY, SPY, by \citealt{Napiwotzki2001}) is compiled from a heterogeneous mix of sources and is tricky to characterize \citep[see][]{Napiwotzki2020}. The second method which is becoming increasingly more utilized recently relies on the synergy between the wide-angle variability surveys such as {\it Kepler} \citep[][]{Howell2014} and the Zwicky Transient Facility \citep[ZTF,][]{Graham2019,Bellm2019}, and {\it Gaia} \citep[][]{Perryman2001,GaiaDR2}. By inspecting the ZTF multi-epoch photometry of WDs selected from the Hertzsprung–Russell diagram the presence of close WD companions is established due to the characteristic narrow eclipses and/or ellipsoidal variations in the lightcurves \citep[see][]{Burdge2019,Burdge2020a,Burdge2020b,Coughlin2020,Keller2021}. In addition to the {\it Gaia}-related selection effects (due to variations of the signal-to-noise of the astrometric measurement), the variability-based detections display a strong period-dependant efficiency evolution \citep[see][]{Korol2017,Burdge2020a,Keller2021}. 

{\it Gaia} offers an opportunity to identify unresolved DWD systems in an entirely different way. Binary orbital motion induces fluctuations of the photocentre position that can be detected with sufficiently accurate astrometry \citep[e.g.][]{Bessel1844}. There are two flavours of the simple unresolved binary detection method available each corresponding to a slightly different orbital separation regime. When modelled with a single-star model at two distinct epochs, binary motion reveals itself as a difference in the measured source proper motion - the effect known as the proper motion anomaly \citep[see e.g.][]{Brandt2018,Kervella2019,Penoyre2020}. Alternatively, the astrometric wobble (and its significance) of the unresolved system's photocentre can be extracted from a single epoch measurement. If the binary period is comparable or shorter than the length of the row of astrometric observations, the single-source model is insufficinet to explain the data thus leading to an excessively large goodness-of-fit statistic. \citet{B20} use the reduced $\chi^2$ of the single-source astrometric fit in the {\it Gaia} DR2 data, also known as Renormalised Unit Weight Error \citep[RUWE][]{RUWE}, to identify binary systems across the entire Hertzsprung–Russell diagram (HRD). They also explain how RUWE can be converted into an estimate of the amplitude of the astrometric wobble, quantity directly comparable to the output of the binary population synthesis models.

\begin{figure*}
  \centering
  \includegraphics[width=0.99\textwidth]{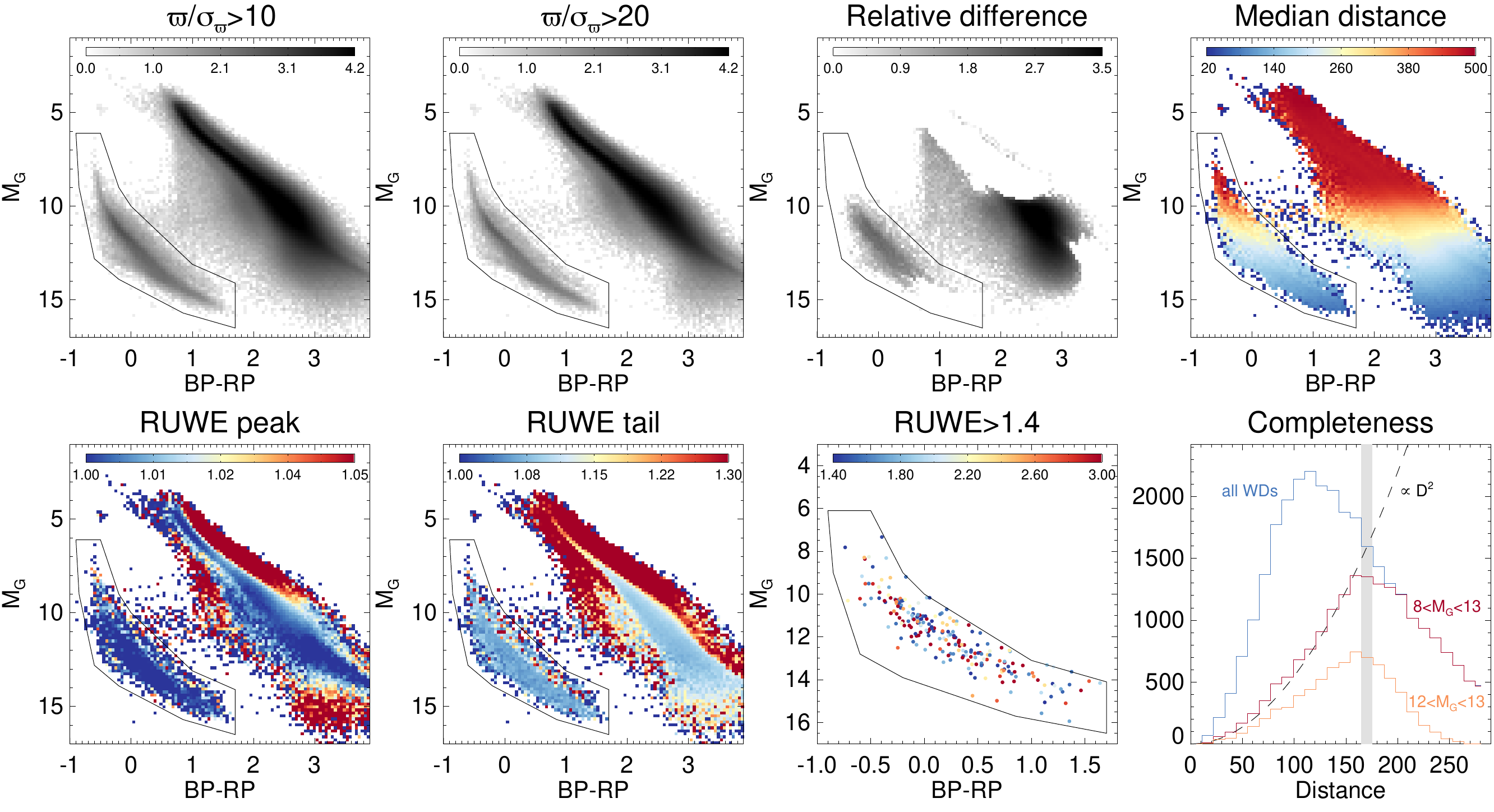}
  \caption[]{Density of {\it Gaia} EDR3 sources on the
    Hertzsprung–Russell diagram (HRD). Only sources i) satisfying the
    first four conditions of Equation~\ref{eq:hrsel} and ii) within
    heliocentric distance of $D<500$ pc are included. {\it Top row
      TR), 1st panel:} shows the HRD behaviour for sources with
    parallax signal-to-noise $\varpi/\sigma_{\varpi}>10$. A large
    cloud of likely spurious objects is clearly seen between the WD
    sequence (outlined in black) and the Main Sequence (MS). While the
    bulk of the spurious stars are limited to $BP-RP>0.7$ (note a
    sharp vertical ``wall'' around this colour) many reach well into
    the WD sequence. {\it TR, 2nd panel:} HRD density for stars with
    $\varpi/\sigma_{\varpi}>20$. Here the WD sequence and the MS are
    well separated, the only connection being the ``bridge'' of likely
    WD-MS binaries at $10<M_G<12$. This is the selection used in our
    analysis. {\it TR, 3rd panel:} relative difference between the two
    density distributions shown in the 1st and 2nd panels. Note the
    excess of faint sources and sources with colours intermediate
    between WD and MS that are most prevalent in the selection with
    the $\varpi/\sigma_{\varpi}>10$ cut. {\it TR, 4th panel:} Distance
    gradient across the HRD. {\it Bottom Row (BR), 1st panel:} HRD
    colour-coded according to the 41st percentile of the RUWE
    distribution, approximately corresponding to the location of the
    peak. The WD sequence is well behaved, but systematic shifts can
    be seen along the MS, e.g. around $M_G\sim7$ and $M_G\sim10$. {\it
      BR, 2nd row:} The colour-coding is according to location of the
    90th percentile of the RUWE distribution (tail). The WD does not
    display a heavy tail. However, a prominent tail is seen along the
    MS photometric binary sequence (just above the MS) and to the blue
    of the MS sequence (possibly spurious astrometry). {\it BR, 3rd
      panel:} WDs with RUWE$>1.4$. {\it BR, 4th panel:} Completeness
    of the WD sample. Distance distribution is shown for the entire WD
    sample (blue) and the sub-sample used for the DWD analysis (red),
    which is complete out to $\sim170$ pc (vertical grey band).}
   \label{fig:selection}
\end{figure*}

\section{Clean sample of White Dwarfs in {\it Gaia} EDR3}
\label{sec:select}

In principle, selecting WDs with {\it Gaia}'s astrometry could not be
easier. Thanks to {\it Gaia}'s accurate parallax measurements, the WD
sequence is easily identified as it sits alone in otherwise
unpopulated portion of the HRD, bluewards and faintwards of the Main
Sequence. However, as revealed by e.g. \citet{GF19} a large number of
spurious sources can enter the region occupied by WDs due to biased
and/or erroneous astrometry. The incidence of DWD binaries is rather
low \citep[see e.g.][]{Toonen2017} and even a percent-level
contamination could therefore significantly bias the measurement we
are after. Below we describe the WD selection procedure we have
adopted to minimise the contamination of our sample by non-WD sources.

In what follows we use the photometric \citep[][]{EDR3_photometry} and
astrometric \citep[][]{EDR3_astrometry} data published as part of the
{\it Gaia} EDR3 \citep[][]{EDR3_summary,EDR3_validation}. We correct
stellar magnitudes and colours for the effects of dust extinction
using the reddening maps of \citet{SFD} and the first two terms for
the extinction coefficients presented for {\it Gaia} DR2 in
\citet{Babu2018}. Note that i) {\it Gaia} EDR3 photometry differs
slightly from that presented in {\it Gaia} DR2 and ii) it would be
more appropriate to use differential (3D) extinction values instead of
those based on total reddening. However, we believe that these
discrepancies have minimal effect on our results given the WD
selection cuts applied (as described below).

The WD sample analysed in this work is procured using the following
criteria:
\begin{equation}
  \begin{aligned}
  &|b|>30^{\circ},\\
  &\verb|PHOT_BP_RP_EXCESS_FACTOR|<3,\\
  &E(B-V)<1,\\
  &13<G<20,\\
  &\varpi/\sigma_{\varpi}>20.
  \end{aligned}
  \label{eq:hrsel}
\end{equation}

\noindent In the list above, Galactic latitude and $BP-RP$ excess flux
cuts aim to minimise the occurrence of blended objects (or sources in
high density environments). Maximal reddening limit helps to ensure
that stars do not wander around the HRD due to unaccounted variations
in differential extinction. The apparent magnitude range reflects the
low intrinsic luminosity of WDs. Finally, the high parallax
signal-to-noise threshold guards against sources with biased/spurious
astrometry.

Figure~\ref{fig:selection} (top row, second panel from the left) shows
the density of sources in the HRD selected using the criteria above
with an additional cut on the heliocentric distance of $D<500$ pc. The
WD sequence is highlighted with a back outline as given in
\citet[][hereafter B20]{B20}. This HRD density can be compared to the distribution
obtained with a lower parallax signal-to-noise threshold (with the
rest of the cuts kept fixed) shown in the first panel of the top
row. Even at a relatively high value of $\varpi/\sigma_{\varpi}>10$ a
large number of spurious detections is visible reaching from the MS
towards the WD sequence. Note that while the bulk of these suspect
sources is limited to $BP-RP>0.7$ (where a sharp vertical boundary is
visible) many percolate bluewards into the WD sequence. The contaminant
nature of these objects is discussed in \citet[][hereafter GF21]{GF21}, see for example
their figure 1. The third panel of the top row of
Figure~\ref{fig:selection} shows a relative density difference between
panel one and two. As expected, the largest discrepancy is at faint
magnitudes, below the MS. However, an excess of stars with lower
astrometric quality can be seen impinging on the WD sequence and
overlapping with it.

In terms of the global properties of the stars selected, the fourth
panel of the Figure's top row gives the median distance behaviour on
the HRD. Unsurprisingly, a strong distance gradient is visible as a
function of absolute magnitude, with stars at $M_G>14$ being limited
to $<100$ pc. 

The first two panels in the bottom row
Figure~\ref{fig:selection} display the details of the distribution of
the astrometric reduced $\chi^2$ as given by RUWE. Namely, the first
panel shows the location of the RUWE peak traced by the 41st
percentile of the RUWE distribution \citep[see][]{RUWE}. By
construction, the 41st percentile of the RUWE distribution should stay
at 1 as the bulk of the stars seen by {\it Gaia} are not expected to
show significant non-linear deviations in the photo-centre motion not
described by the standard model. This assumption can of course be
broken in the regions of the HRD dominated by binaries, such as the
binary MS, where the shape of the RUWE distribution is altered, as
indeed demonstrated by the first panel in the bottom row of
Figure~\ref{fig:selection}. Worryingly however, this panel shows
additional areas in the HRD where deviations of the RUWE peak from the
expected value is visible, for example, red region (strong deviation)
directly underneath the MS and pale blue region (weak deviation) along
the MS, in particular at $M_G\sim7$ and $M_G\sim10$. Reassuringly, the
WD sequence is almost uniformly lit up with a dark blue colour,
indicating that peak of the RUWE distribution remains very close to
1. Second panel in the bottom row of Figure~\ref{fig:selection}
illustrates a surprisingly heavy tail in the RUWE distribution (note the difference in the RUWE range displayed)
in some portions of the HRD, in particular above and below the MS. Again, across the WD sequence the RUWE tail is not pronounced
and remains approximately constant for WDs of all colours and
magnitudes.

The next (third) panel in the row gives positions of the
WDs with a clear RUWE excess, which is defined as RUWE $>1.4$. These systems are likely unresolved binaries. Their distribution in the HRD is
understandable: they congregate to the regions of the highest WD
density (see top left panel of the Figure), with many objects sitting
slightly above this central ridge due to the addition of the fluxes of
the individual components. The rightmost (fourth and final) panel of
the bottom row shows luminosity functions (LF) of several WD
sub-samples. For example, the light-blue histogram presents the LF of
the whole sample. The LF is peaked around 100 pc betraying
incompleteness of the sample beyond this distance \citep[see also][for
  discussion of the WD sample completeness]{Rix2021}. While
intrinsically bright WDs are seen out to large distances, e.g. $D>500$
pc, the dimmer (and more numerous) stars are only detected out to 100
pc or so. To choose an optimal absolute magnitude cut we consider the
following three effects: completeness, contamination and astrometric
wobble sensitivity. Figure~\ref{fig:gf_comp} demonstrates that the
contamination of the sample with (likely) spurious objects grows at
$M_G>13$.  As discussed in \citetalias{B20} and as we show below,
the sensitivity of the astrometric wobble drops quickly with
distance. According to Figure~\ref{fig:dwd_meas}, the background of
sources with no significant astrometric $\chi^2$ excess starts to
overwhelm the signal at heliocentric distances $D>170$ pc. We
therefore limit our WD selection to $8<M_G<13$ as this sample is
complete to $D=170$ pc (red histogram in the fourth panel of
Figure~\ref{fig:selection}). Note that even the sample of the faintest
sources ($12<M_G<13$) is complete out to this distance (orange
histogram).

\subsection{Comparison with GF21}

Most recently, \citetalias{GF21} presented a catalogue of WDs
identified in the {\it Gaia} EDR3 data. Their training set is based on
spectroscopic WDs which allows them to produce a clean and robust
selection of WDs. Unfortunately, we can not use their WD sample to
measure the incidence of DWD binaries as \citetalias{GF21} included
criteria based directly on the quality of the {\it Gaia} astrometric
solution (for example RUWE, excess astrometric noise, etc), thus
biasing their sample against systems with significant astrometric
wobble. Nonetheless, we will compare the make-up of our WD catalogue
against that of \citetalias{GF21} with the goal to identify regimes
free of obviously spurious contaminants. 

\begin{figure}
  \centering
  \includegraphics[width=0.48\textwidth]{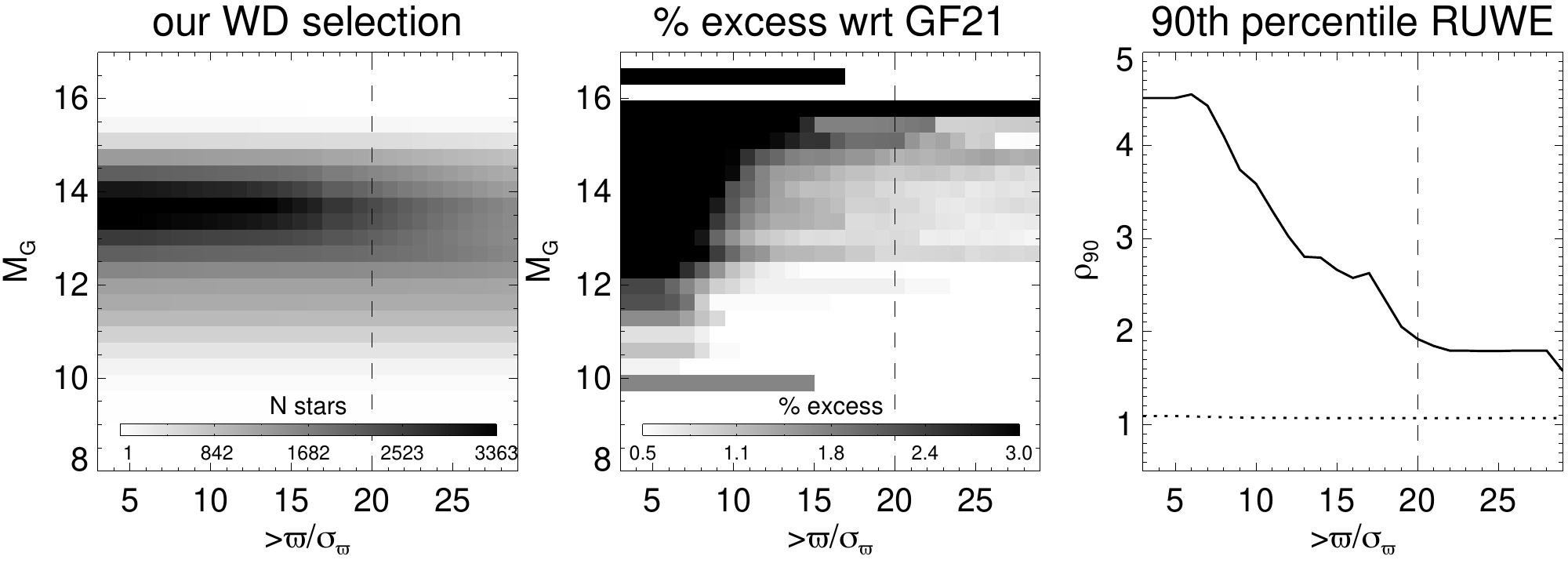}
  \includegraphics[width=0.48\textwidth]{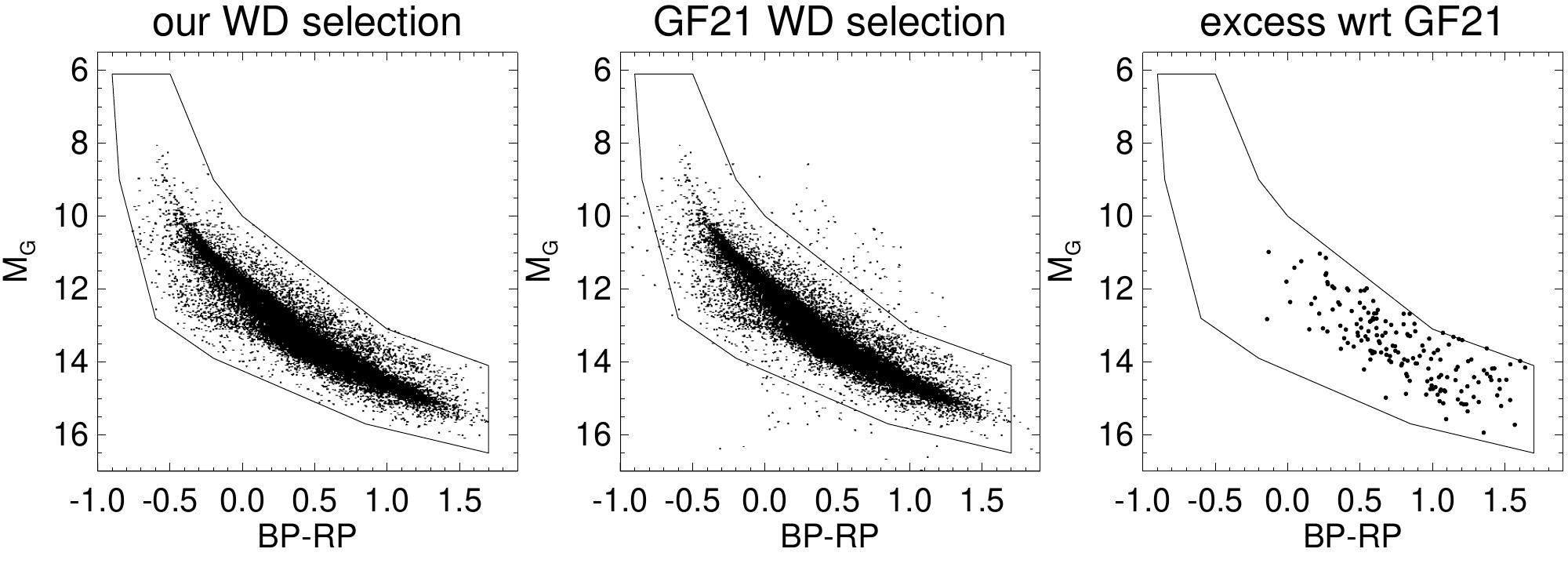}
  \caption[]{Comparison between our selection of WDs
    (Equation~\ref{eq:hrsel} plus HRD mask) and that of
    \citetalias{GF21}. Note that for \citetalias{GF21} sample we
    additionally apply cuts from Equation~\ref{eq:hrsel}. {\it Top Row
      (TR), 1st panel:} number of WDs as a function of magnitude $M_G$
    ($y$-axis) and parallax signal-to-noise $\varpi/\sigma_{\varpi}$
    ($x$-axis). {\it TR, 2nd panel:} Excess of WD stars in our
    selection with respect to \citetalias{GF21} as a function of
    magnitude $M_G$ ($y$-axis) and parallax signal-to-noise
    $\varpi/\sigma_{\varpi}$ ($x$-axis). {\it TR, 3rd panel:} 90th
    percentile of the RUWE distribution for the stars in common
    between our selection and that of \citetalias{GF21} (dotted line)
    and the excess of stars in our selection (solid) as a function of
    parallax signal-to-noise $\varpi/\sigma_{\varpi}$. Note the
    prominent RUWE tail in the excess which decreases steadily with
    $\varpi/\sigma_{\varpi}$ but stabilizes for
    $\varpi/\sigma_{\varpi}>20$. {\it Bottom row (BR), 1st panel:}
    Distribution of WDs in our selection (given by
    Equation~\ref{eq:hrsel}). {\it BR, 2nd panel:} Distribution of WDs
    in \citetalias{GF21} (together with the cuts from
    Equation~\ref{eq:hrsel}). {\it BR, 3rd panel:} WDs included in our
    selection but not present in \citetalias{GF21}.}
   \label{fig:gf_comp}
\end{figure}

As discussed above and shown in the top row of Figure~\ref{fig:selection}, the contamination of the
WD sample is a function of the parallax signal-to-noise
ratio. Accordingly, top row of Figure~\ref{fig:gf_comp} shows the
properties of our WD selection as a function of
$\varpi/\sigma_{\varpi}$ cut. The top left panel gives the evolution
of the WD LF as a function of the parallax S/N. The middle panel in
the top row shows the percentage excess in our WD samples compare to
that in \citetalias{GF21}. The excess shows a clear evolution with
$\varpi/\sigma_{\varpi}$, dropping significantly for
$\varpi/\sigma_{\varpi}>10$. But even for $\varpi/\sigma_{\varpi}>10$,
the excess can be of order of percent(s), in particular for faint
systems with $M_G>13$. The evolution of the excess of WDs in
comparison to the \citetalias{GF21} can be compared with the evolution
of the tail of the RUWE distribution of the stars in common between
the two samples (dotted line) and stars that are part of the excess
(i.e. present in our selection but not identified as WD by
\citetalias{GF21}) as a function of $\varpi/\sigma_{\varpi}$ (third
panel in the top row of the Figure). The RUWE tail (as estimated by
the 90th percentile of the RUWE distribution) for the stars in common
is not evolving as a function of the parallax S/N, hovering just above
RUWE$=1$. This can be compared to the behaviour of the RUWE tail in
the excess which is continuously decreasing from worryingly high
values for $\varpi/\sigma_{\varpi}<20$, but beyond that, stays flat at
just below RUWE$=2$. Finally, the bottom row of
Figure~\ref{fig:gf_comp} compares the HRDs positions of WD in our
sample (i.e. selected according to Equation~\ref{eq:hrsel}, left) and
\citetalias{GF21} (i.e. present in \citetalias{GF21} together with
cuts from Equation~\ref{eq:hrsel}, middle). Note that the
\citetalias{GF21} contains a small number of WDs outside of the main
WD sequence (outlined by black line). Most of these systems are
binaries in which WDs are paired with non-degenerate companion
(e.g. WD plus M dwarf). The third (and final) panel of the bottom row
of the Figure shows the WDs (i.e. stars located within the black WD
mask) that are present in our catalogue but absent in
\citetalias{GF21}. The bulk of the excess is at faint magnitudes with
only a small number of systems brighter than $M_G=13$.

\section{DWD fraction from astrometry}
\label{sec:measure}

\citetalias{B20} presented a strategy to estimate the fraction of
unresolved binaries using the amplitude of the positional wobble
obtained from the reduced $\chi^2$ of the astrometric model for each
source. In this Section, we briefly discuss the steps of the procedure
and add several modifications.

\subsection{Measurement procedure}\label{sec:mes}

{\it Rescaling of RUWE}. Before we proceed to the calculation of the
wobble, we choose to rescale the published RUWE values to minimise the
influence of the spurious shifts of the peak of the RUWE distribution
(see bottom row of Figure~\ref{fig:selection}). This is done for
subsets of the WDs in our sample, in bins of heliocentric
distance. More precisely, for each distance bin we re-scale RUWE so
that the 41st percentile of its distribution is precisely 1. Note that
for the WD sample under consideration, this rescaling is minimal as
the 41st percentile variation (before rescaling) across the entire
distance range of 0-170 pc is limited to between 0.988 and 1.007. We
have checked that the measurements of the DWD fraction reported below
are not affected by the rescaling.

\begin{figure*}
  \centering
  \includegraphics[width=0.99\textwidth]{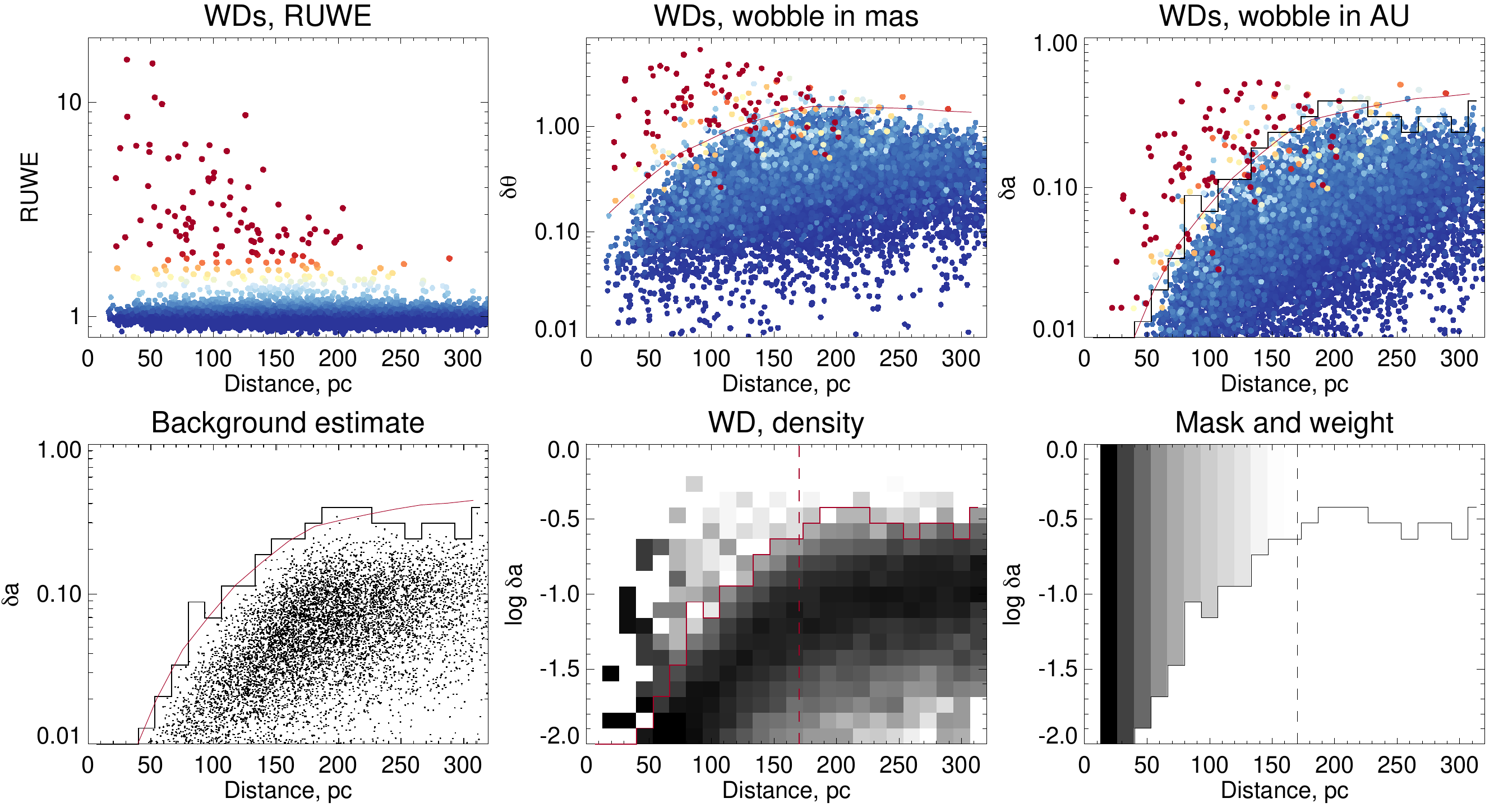}
  \caption[]{Astrometric wobble as a function of distance. {\it Top
      row (TR), 1st panel:} RUWE of the WDs selected using
    Equation~\ref{eq:hrsel} and the HRD mask shown in
    Figures~\ref{fig:selection} (and ~\ref{fig:gf_comp}). {\it TR, 2nd
      panel:} Angular wobble amplitude $\delta\theta$ as a function of
    distance; colour-coding is according to RUWE as shown in the
    previous panel. Note that while the signal drops with distance
    (yellow/red points) the background grows and starts to dominate
    beyond $D\sim170$ pc. {\it TR, 3rd panel:} Physical wobble $\delta
    a$ as a function of distance, same colour-coding. Black histogram
    and red curve give two estimates of the boundary between
    background-dominated and background-free regimes (see below and
    main text for details). {\it Bottom row (BR), 1st panel:} RUWE
    values for a model ``background'' population of source
    (i.e. single stars or source with insignificant wobble) obtained
    by reflecting the RUWE distribution about the peak (small black
    dots). Black histogram shows the value of RUWE above which no
    background sources exist at given distance. A different estimate
    of the location of the background-free regime is given by the red
    curve (obtained using $RUWE=1.4$, see main text for details). The
    two estimates are in good agreement. {\it BR, 2nd panel:} Density
    of observed sources in the plane spanned by $\delta a$ and
    distance. Note a dramatic switch in the density behaviour above
    the red line (i.e. on moving into background-free regime). {\it
      BR, 3rd panel:} Weight (greyscale) and mask used to convert
    detections of DWD at different distances into a single DWD
    estimate (see main text for details).}
   \label{fig:dwd_meas}
\end{figure*}

{\it Converting RUWE into wobble amplitude $\delta a$}. Following \citetalias{B20} we
convert {\it Gaia} EDR3 RUWE values into the amplitude of the angular
wobble $\delta\theta$ using their Equation 2, namely $\delta\theta \approx \sigma_{\rm AL}(G)~\sqrt{\rho^2-1}$. This requires
multiplication of RUWE $\rho$ by the appropriate uncertainty of the centroid
measurements (in the along-scan direction) $\sigma_{\rm AL}$ for which
we use the robust scatter estimate as given by the solid blue line in
Figure A.1 of \citet{EDR3_astrometry}. Note that our $\sigma_{\rm AL}$
estimate is dependent only on the source magnitude $M_G$. The angular
wobble amplitude $\delta\theta$ is then converted into the
corresponding displacement $\delta a$ in au using Equation 4 of
\citetalias{B20}. Measured $\delta a\equiv a\delta_{ql}$ is related to the binary
separation $a$ via the scaling factor
$\delta_{ql}=|q-l|(q+1)^{-1}(l+1)^{-1}$, where $q$ and $l$ are the
mass and luminosity ratio correspondingly.

{\it Strong wobble vs weak wobble}. As argued in \citetalias{B20}, the
stellar population's binary fraction ought to be estimated for a fixed
$\delta a$ threshold across the distance range probed. However, at
fixed $\delta a$ the contribution of the background of sources with no
significant wobble increases with distance. To take the background
contribution into account, \citetalias{B20} built a model of the RUWE
distribution of well-behaved and/or single sources by reflecting the
measured RUWE distribution around the peak (essentially using objects
with RUWE $<1$ to generate mock sources with RUWE $>1$ assuming the
symmetry of the RUWE distribution around the peak). An example of the
behaviour of such a background model in the space of $\delta a$ and
distance is shown in the second panel of Figure 8 in
\citetalias{B20}. With the background model in hand, \citetalias{B20}
calculated the excess of objects in the range of $\delta a$ and
distance (see third panel of their Figure 8). They argued that the
range of $\delta a$ must be chosen to lie between the two regimes,
that dominated by the background at low $\delta a$ and that of
``resolved'' systems at high $\delta a$. While it is true that at very
high $\delta a$ values (corresponding to separations of more than
$\approx0.1$ arcsecond) binary systems would be resolved by {\it
  Gaia}, a good portion deemed by \citetalias{B20} as resolved are
actually unresolved binaries with {\it strong} astrometric
wobble. These are the systems with RUWE values far in the tail of the
RUWE distribution, i.e. likely with RUWE~$>1.4$. This is confirmed by
the red colour (strong excess compared to the background) of the
region designated as ``resolved'' in Figure 8 of \citetalias{B20} -
the background population does not contribute sources with high
$\delta a$ values at these small distances. Rather than being
resolved, these objects are in the {\it strong wobble} regime to be
compared with the {\it weak wobble} displayed by the stars used by
\citetalias{B20}. Here we refer to the {\it weak wobble} stars as
those in the main peak of the RUWE distribution (typically with
$1<$RUWE$<1.4$) and {\it strong wobble} stars as those in the tail of
the RUWE distribution (typically, with RUWE$>1.4$). Binary estimates
based on the weak and strong wobble systems should in principle
agree. However, the fidelity of the background model plays a much more
important role in the case of the weak wobble estimate. In the strong
wobble regime, true unresolved binaries dominate and the contribution
of the single and/or well-behaved objects is essentially zero. In what
follows, we measure the DWD binary fraction relying solely on systems
in the strong wobble regime. 

Figure~\ref{fig:dwd_meas} details the steps of the DWD measurement
process. Left panel in the top row shows the evolution of RUWE values
for WD selected using Equation~\ref{eq:hrsel} as a function of
distance. As expected for genuine binaries, objects with the highest
values of RUWE congregate at small distances and the overall density
of RUWE is to drop with increasing distance. Next (middle) panel in
the same row shows the amplitude of angular wobble $\delta\theta$ as a
function of distance. The $y$ axis is logarithmic and the
colour-coding of symbols is according to the RUWE value as shown in
the previous panel. For genuine binaries the angular wobble amplitude
decreases (on average) with distance (see the upper envelope of the
distribution, in particular for red symbols). For the background
(i.e. for single and/or well-behaved sources) it instead increases,
this is because the width of the peak of the RUWE distribution stays
more or less the same but the typical apparent magnitude of the source
increases, and, therefore, so does $\delta\theta$. The right (and
final) panel in the top row of Figure~\ref{fig:dwd_meas} shows the
amplitude of the astrometric wobble $\delta a$ as a function of
distance. Here, decreasing angular sensitivity of {\it Gaia} and the
distance scaling of $\delta a$ cancel out and the upper envelope of
the distribution is almost flat. Multiplication by distance increases
the background which starts to dominate beyond $D=170$~pc. In this and
the previous panel, a simple estimate of the boundary between the
strong and weak wobble regimes is shown as a red curve. This boundary
is obtained by converting the critical value of RUWE$_{\rm crit}=1.4$
into $\delta\theta_{\rm crit}(D)$ by using the astrometric error
$\sigma^{\rm med}_{\rm AL}(D)$ corresponding to the median apparent magnitude in
each distance bin.  Values of critical $\delta a_{\rm crit}(D)$ are
procured by simply multiplying $\delta\theta_{\rm crit}(D)$ by the
corresponding distance of each bin.

Left panel in the bottom row of Figure~\ref{fig:dwd_meas} presents a
different method of estimating the background. In this case, we follow
the prescription of \citetalias{B20}: using sources with RUWE$<1$, we
generate mock objects with -RUWE$+2$ (assuming that the peak is at
RUWE $=1$), i.e. `reflecting' the RUWE distribution around the
peak. These new RUWE values are converted into $\delta\theta$ and
$\delta a$ using Equations 2 and 4 of \citetalias{B20}. Black
histogram in the left panel of the bottom row of the Figure gives the
threshold value of $\delta a^{\prime}_{\rm crit}(D)$ above which no
mock background sources are found. As expected, the two estimates of
the location of the background-free regime $\delta a_{\rm crit}(D)$ (shown as a solid red curve in the bottom left panel)
and $\delta a^{\prime}_{\rm crit}(D)$ (given as a histogram in the bottom row) are in good agreement. The
middle panel of the bottom row in Figure~\ref{fig:dwd_meas} displays
the density of observed sources in the plane of $\delta a$ and
distance. Note the dramatic change in the behaviour of the density on
transition from the background-dominated into the background-free
regime. As illustrated in the Figure, the number of identifiable
unresolved binaries is dictated by {\it Gaia}'s sensitivity to
photo-centre wobbling which is a strong function of distance,
convolved with the number of sources detected that grows with the
volume probed. At small distances, e.g. $D<50$ pc, the entire range of
$10^{-2}<\delta a<1$ is accessible but the volume is small and so is
the total number of binaries. At larger distances, e.g. $D>100$ pc,
the increase in the volume brings in a large number of wobbling WDs,
but only systems with large amplitudes can be caught, e.g. those with
$\delta a>0.1$.

To produce an estimate of the DWD binary incidence, we combine
detections of the DWDs from the entire distance range ($0<D/\text{pc}<170$)
by scaling contributions from each distance bin proportionally to the
inverse of the number of WDs detected (equivalent to the volume
probed). In practice, the DWD number from each bin is rescaled to the
total number of WDs in 170 pc$^3$ volume. Bottom right panel of
Figure~\ref{fig:dwd_meas} shows the mask used to select only the
contributions from the background-free regime as well as the weight
assigned to each distance bin (greyscale). There are 8568 WDs in our
sample within 170 pc. We estimate that there are 555 DWD systems with
$10^{-2}<\delta a<1$, based on 157 DWD detections. The number of
systems with RUWE $>1.4$ is slightly smaller, i.e. 118 but switching to
this sub-sample does not change our estimate of the DWD fraction.

\subsection{DWD wobble amplitude distribution}
\label{sec:wa_dist}

\begin{figure}
  \centering
  \includegraphics[width=0.48\textwidth]{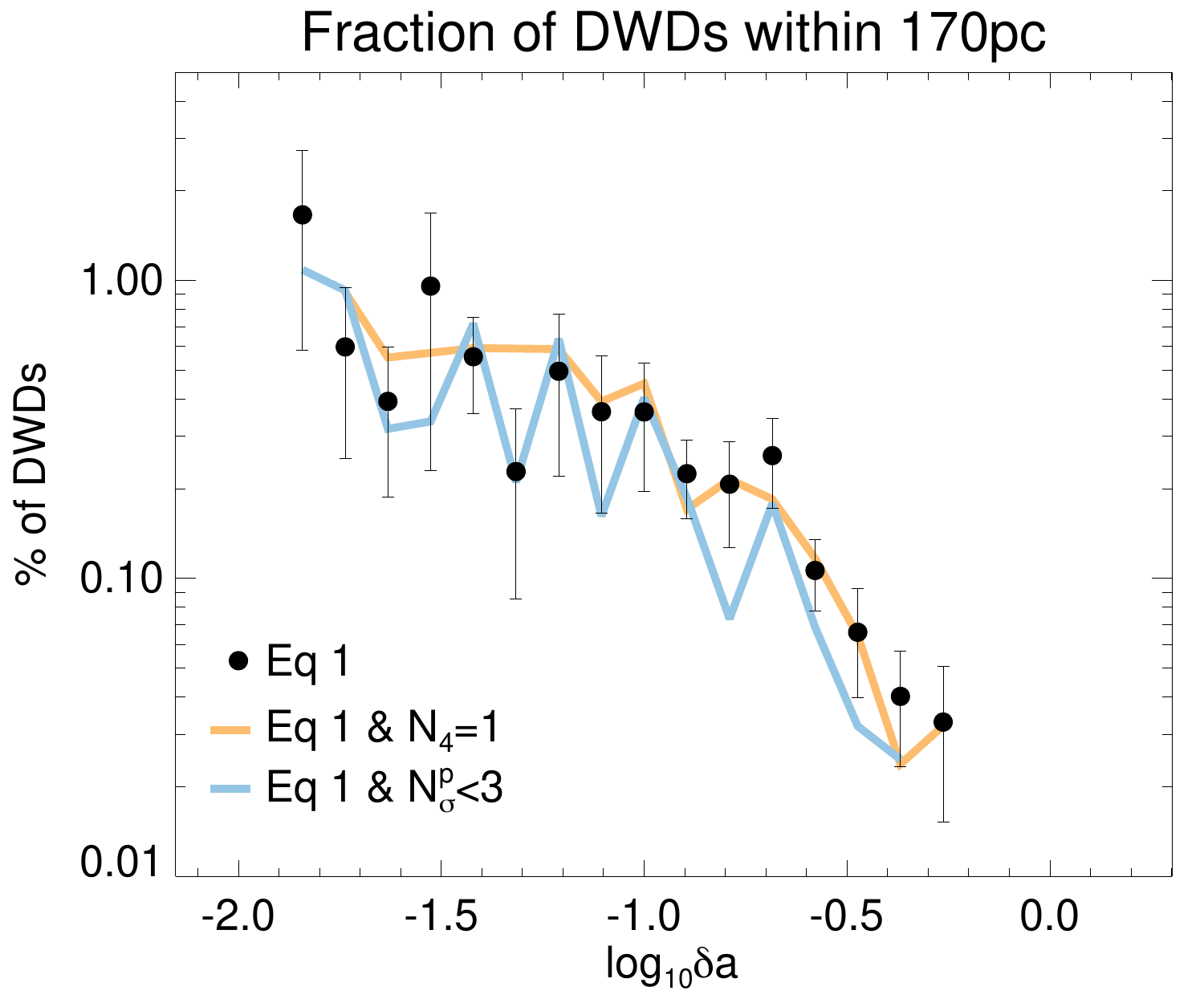}
  \caption[]{Fraction of DWDs within 170 pc from the Sun. Filled-in
    circles with error-bars show the DWD fraction as a function of
    logarithm of the wobble amplitude $\log\delta a$ for the main
    sample. Light orange (blue) curve represent the DWD fraction
    measurement for a sample of WDs with no other {\it Gaia} EDR3
    detections within 4 arcseconds (with low photometric variability
    as given by $N^{p}_{\sigma}<3$, see main text for details). DWD
    fraction estimates for all three samples agree. Note a sharp drop
    of the DWD fraction for systems with $\log_{10}\delta a>-1$.}
   \label{fig:dwd_da}
\end{figure}

 Figure~\ref{fig:dwd_da} presents the probability
distribution of $N(\delta a)$ of the wobble amplitude in our
sample. There are two properties of the DWD p.d.f. worth noting: i)
the rise towards low wobble amplitudes and ii) the precipitous drop at
$\log_{10}\delta a>-0.7$. The prevalence of the low $\delta a$ is
already hinted at in the right panel of the top row of
Figure~\ref{fig:dwd_meas}. Here, there is a clear lack of systems with
high wobble amplitudes, e.g. those with $\delta a> 0.1$ at low
distances ($D<50$~pc); these systems only begin to be sampled when a
large enough volume of the Galaxy is probed. 

To investigate whether the DWD fraction estimates could be affected by the presence of spurious contaminants we study two additional sub-samples. The first set systems is limited to the WDs with no {\it Gaia} EDR3 neighbours
within the radius of 4 arcseconds $N_4=1$. The wobble amplitude
p.d.f. for this sub-sample is shown in Figure~\ref{fig:dwd_da} as
light orange curve in complete agreement with the measurement based on
the larger sample. For our second sub-sample we limit the WD selection
to systems with low photometric variability. We use the {\it Gaia}
EDR3 flux error to gauge the photometric variability amplitude
following the methodology of \citet{Belokurov2017}. To account for the
gradual increase of the flux error with apparent magnitude, we
calculate the median and standard deviation of the photometric
amplitude in bins of magnitude $G$ and calculate for each star its
excursion from the median in terms of the number of standard
deviations $N^{p}_{\sigma}$. The DWD fraction estimate using this
sample is shown with light blue curve in the Figure. The later is also in
good agreement with the DWD fraction measurement based on the entire
WD sample.

\subsection{Double WD vs double MS systems}

As a sanity check, let us contrast the distribution of the derived
wobble amplitudes for DWDs to that of the double MS (DMS) systems. For
comparison, we will use the stars within the same absolute magnitude
range $8<M_G<13$ but located on the MS (using the MS mask as presented
in \citetalias{B20}). These stars are similar to the WDs studied in
terms of their observed properties, for example, their apparent
magnitudes are similar and therefore so are the astrometric errors;
given the identical absolute magnitude range, they probe similar
distances from the Sun. However, the unresolved MS binaries contained
within the sample have not yet had a chance to evolve (e.g. become red
giants, go through a CE phase, interact etc) and
therefore are expected to have a drastically different period (and
consequently, separation) distribution. More precisely, the period
distribution of low-mass MS stars has been shown to be broad and is
estimated to peak at around $10^5$ days \citep[see
  e.g.][]{Raghavan2010}. 

\begin{figure}
  \centering
   \includegraphics[width=0.48\textwidth]{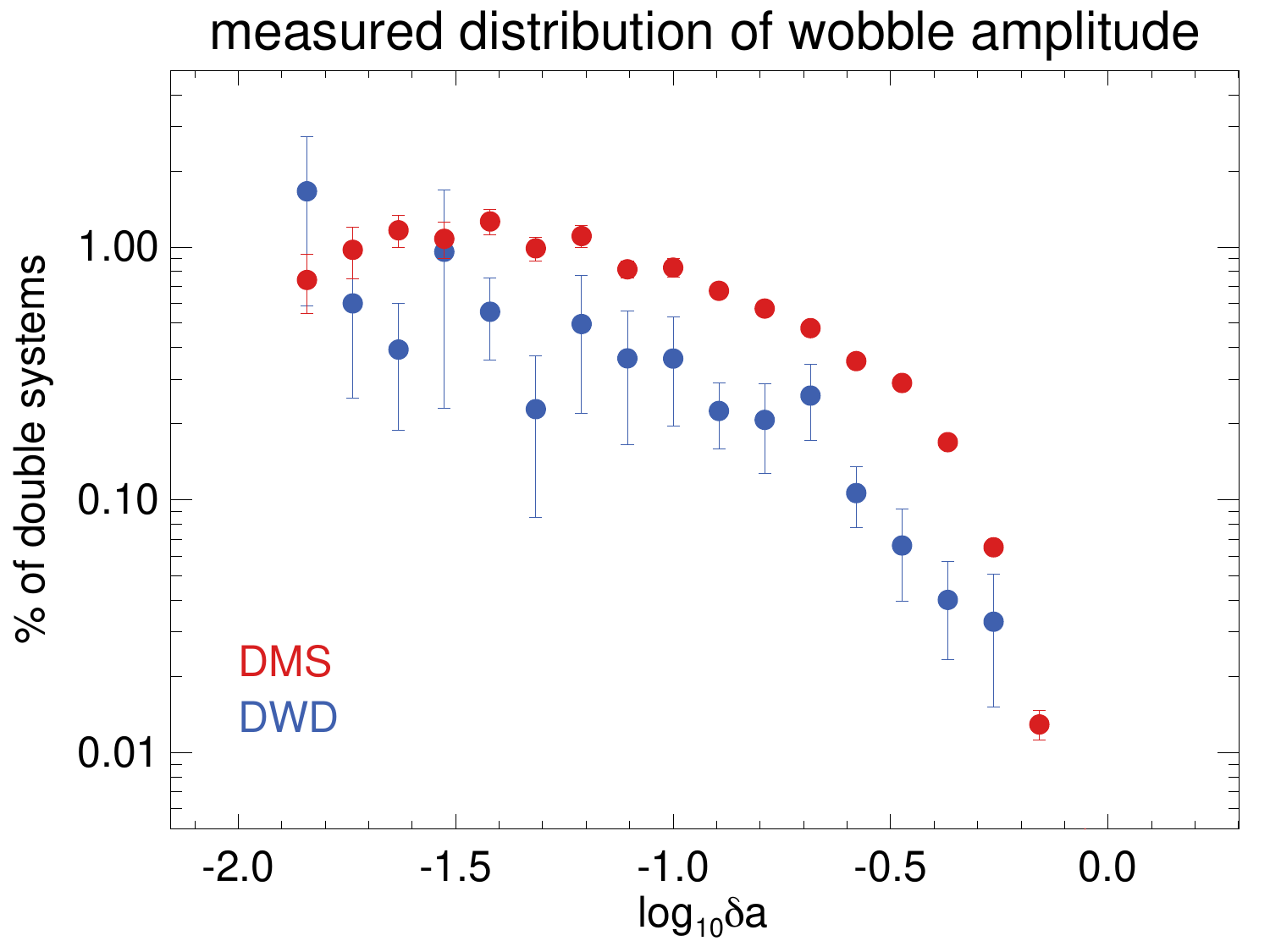}
  \includegraphics[width=0.48\textwidth]{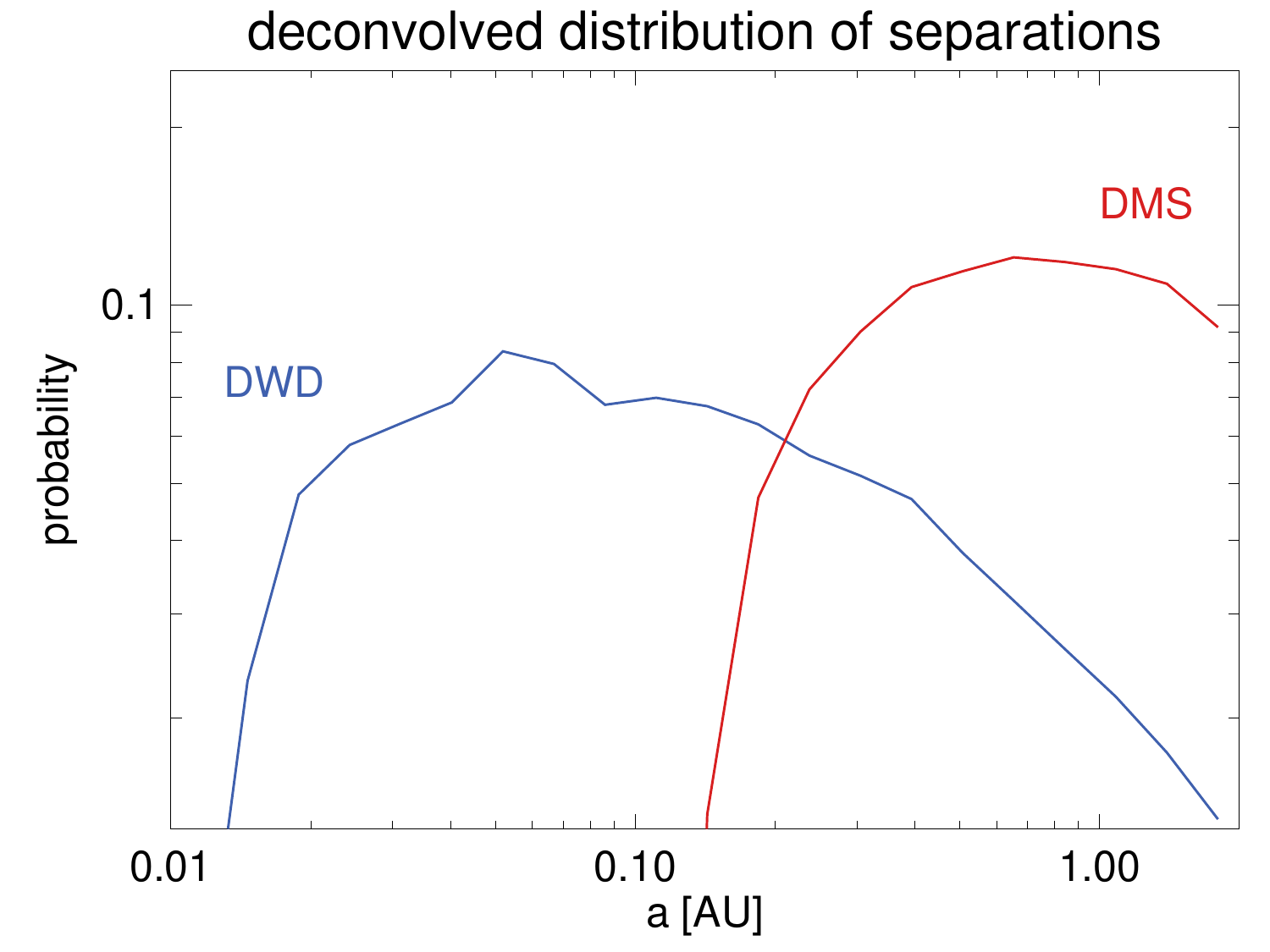}
  \caption[]{{\it Top:} Distribution of wobble amplitude $\delta a$
    for DMSs (blue) and DWDs (red). {\it Bottom:}`Deconvolved'
    separations distributions, i.e. each $\delta a$ contributing to
    the distributions shown in the top panel is divided by
    $\delta_{ql}$ value randomly drawn from the distributions shown in
    Figure~\ref{fig:dql}.}
   \label{fig:dwd_dms}
\end{figure}

We select the MS stars using
conditions listed in Equation~\ref{eq:hrsel} and we repeat the
sequence of steps described in Section~\ref{sec:mes}. Note that while
for the WD population the peak of the RUWE distribution is close to 1
in the original {\it Gaia} EDR3 data, for the MS we detected
noticeable shifts (see the bottom row of Figure~\ref{fig:selection})
and therefore RUWE rescaling plays a much more important role. Top
panel of Figure~\ref{fig:dwd_dms} gives the fraction of DMS stars as a
function of wobble amplitude $\delta a$ (red filled circles). Across
the entire range of $\delta a$ probed, these fractions are higher than
those for DWDs (blue filled circles). Note that in the Figure the two
distributions are compared on logarithmic scale, and at some values of
$\delta a$, the incidence rate of DMS binaries exceeds that of DWDs by
are almost an order of magnitude more frequent than double WDs. To
illustrate the differences in the underlying separation distributions
of the two binary samples, we deconvolve their $\delta a$
distributions by dividing each $\delta a$ value by a $\delta_{ql}$
randomly drawn from a model distributions as discussed in Appendix~\ref{ref:dwd_dms_detail}.

The the resulting `deconvolved' distributions of binary separations are
shown in the bottom panel of Figure~\ref{fig:dwd_dms}. As anticipated,
the shapes of the DWD and DMS separation distributions are utterly
different. The DWDs prefer small separations with the majority of
systems at $a<0.1$\,au and a marked drop of systems with larger
separations. On the other hand, there is paucity of DMSs with $a<0.1$\,au with the incidence of binaries growing at higher separations. Note
that both distributions are affected by selection biases. {\it Gaia}
EDR3 astrometry struggles to pick up systems with a wobble amplitude
$\delta a\leq10^{-2}$, as the typical astrometric error surpasses the
size of the photo-centre wobble. Given the distribution of the
$\delta_{ql}$ (in particular, the long tail towards low
$\delta_{ql}$ (cf. Figure~\ref{fig:dql}), this implies that many systems will be missed at separations approaching $a\approx10^{-2}$\,au. This is reflected in the turn-over and a drop in the separation distribution of DWD at $a<0.04$\,au. With this in mind, we conjecture that the intrinsic DWD
distribution likely continues to rise steadily below $a=0.01$\,au. A
different selection effect takes place at larger separations,
i.e. those of $\approx$1\,au or thereabouts. For binary systems with
periods close to the time span of {\it Gaia} EDR3, some of the
photo-centre shift will be absorbed by the proper motion measurement
as the portion of the orbit traced becomes quasi-linear (thus
producing a proper motion anomaly signature). This is discussed in
greater detail in \citet{Penoyre2021}, see their Figures 15-17. Indeed,
given the typical combined mass of the DMS binary of $\approx0.5$\,M$_{\odot}$, there is a large fraction of systems with periods above 2 years (roughly the timespan of {\it Gaia} EDR3) at separation of 1\,au
and above. This explains the flattening and the turn-over of the DMS
separations in this regime. The intrinsic distribution of separations
likely continues to rise. The typical combined mass of DWDs is larger,
at $\approx 2$\,M$_{\odot}$, and therefore their periods are typically
shorter. Thus, for DWDs (as compared to DMS systems within the same
absolute magnitude range) the transfer of power from the RUWE
signature into proper motion anomaly happens at larger separations,
i.e. $a>2$\,au.

\section{Synthetic models of the Galactic DWD population} \label{sec:bps}

We construct synthetic models of the local DWD population using binary population synthesis (BPS) code {\sc SeBa} \citep{PZ96,Nelemans01a,Toonen12}. The detailed description of the considered models is presented in \citet{Toonen12,Toonen13,Toonen2017} to which we refer for further details, while here we summarise their most important characteristics.

\subsection{Initial population} \label{sec:initial_distributions}

The initial stellar Zero Age Main Sequence (ZAMS) population is obtained with a Monte Carlo approach, assuming Solar metalicity, the initial binary fraction of 50\, per cent, and distributions for the binary parameters as detailed below. The mass of the primary stars is drawn between 0.95\,M$_\odot$ - 10\,M$_\odot$ according to the initial mass function of \citet{KroupaIMF}. The mass of the secondary star is drawn from a flat mass ratio distribution between 0 and 1 \citep[e.g.][]{Duchene2013}.
Semi-major axes are drawn from a distribution that is uniform in $\log(a)$ (Abt 1983). We consider only detached binaries on the ZAMS with orbital separations up to $10^6\,$R$_\odot$, and for the orbit eccentricities we assume a thermal distribution \citep{Heggie1975,Raghavan2010,Duchene2013}. 

\subsection{Binary evolution} \label{sec:DWD_evolution}

We let {\sc SeBa} evolve the initial population until both stars turn into WDs. Typically, two stars that start on wide orbits of $\gtrsim$10\,au, in {\sc SeBa} end up evolving independently, as they would if they were born in isolation. However, two stars starting on a short orbit are destined to affect each other via a number of processes such as mass and  angular momentum transfer, CE evolution, magnetic braking, and gravitational radiation. These and other processes typically involved in the binary evolution are modeled within {\sc SeBa} with appropriate recipes detailed in \citet[][and references therein]{PZ96, Toonen12}. Below we describe those most relevant for this study.

\subsubsection{Common envelope evolution} \label{sec:CE}

The CE evolution is a short phase lasting up to thousands of years that occurs when one star of the pair expands and engulfs the companion \citep[see][for a review]{Ivanova2013}.
The engulfed star therefore moves through the companion's envelope experiencing dynamical friction, which facilitates the transfer of the binary orbital energy and angular momentum to the envelope.
Typically, this process is implemented in BPS codes by parametrising either the energy balance equation through an $\alpha$ parameter or that of the angular momentum through a $\gamma$ parameter \citep{Paczynski1976,Webbink1984, Liv88, Nelemans2000}.
Both parameters, $\alpha$ and $\gamma$, encode the efficiency with which orbital energy and/or angular momentum are consumed for unbinding of the envelope and for shrinking of the binary. We highlight that the $\gamma$-parametrisation was originally formulated and fine-tuned based on the reconstructed evolutionary paths of the observed DWDs \citep{Nelemans2000, Nelemans2005, van06}. Models utilised in this study are based on the combination of these two CE formalisms. In addition, different CE efficiencies are considered (cf. Section~\ref{sec:CEefficiency}).

Following \citet{Toonen12} we generate two families of models, which we denote $\alpha \alpha$  and $\gamma \alpha$. In the $\alpha \alpha$ model, the $\alpha$-formalism is applied to determine the outcome of every CE phase. In the $\gamma \alpha$ model, the $\gamma$-prescription is applied unless the binary contains a compact object or the CE is triggered by a tidal instability, in which case $\alpha$-prescription is used.
Thus, in the $\gamma \alpha$ evolution model the first CE phase is typically described by the $\gamma$ formalism, while the second by the $\alpha$ formalism.  It is worth noticing that the treatment of CE evolution has an effect on the DWD formation rate. In particular, we note that the $\gamma \alpha$-model yields about twice as many DWDs compared to the $\alpha \alpha$-model \citep[e.g.][]{Toonen2017}.  

\subsubsection{Common envelope efficiency}
\label{sec:CEefficiency}
In the standard {\sc SeBa} setting $\gamma = 1.75$ and
$\alpha \lambda =2$ (here $\lambda$ is a parameter dependent on the structure of the donor star), both values are derived based on the observed DWD sample \citep{Nelemans2000,Nelemans2005}. However, lower CE efficiencies have been found to fit the population of the post-common-envelope binaries (i.e. WD+MS, \citealt{Zorotovic2010, Toonen13, Camacho2014}).
Therefore, to complement models using the standard {\sc SeBa} efficiency values $\alpha \alpha$ and $\gamma \alpha$, we construct two additional model variations $\alpha \alpha$-2 and $\gamma \alpha$-2 setting $\alpha \lambda = 0.25$. 
In these models the binary orbit shrinks more strongly compared to the standard efficiency value because for lower values of $\alpha \lambda$ the envelope is expelled less efficiently.
The effect of a stronger orbital decay is reflected in the number of DWD systems produced by these models:
the $\alpha \alpha$-2 model generates 7 times fewer binaries with $a < 1\,$au compared to the default $\alpha \alpha$-model because the contraction of the orbit during the CE evolution is so strong that many do not survive until the DWD stage (cf. Table~\ref{tab:models}).

\subsubsection{Stable mass-transfer } \label{sec:SMT}

When one of the two stars fills its Roche lobe,
the subsequent flow of matter may be self-stabilising, unlike in the previously discussed case leading to a CE-phase. 
In our simulations the stability and the rate of mass-transfer are dependent on the reaction  of the stellar radii and the Roche lobes to the transfer (and possible loss) of mass and angular momentum. 
The loss of angular momentum is poorly constrained by observations. Consequently, binary evolution codes make different assumptions \citep[for comparison see][]{Toonen14}. Often the orbital angular momentum is set to leave the binary with (a multiple of) the specific orbital angular momentum of the binary $\beta=$ constant.
In {\sc SeBa}, for a non-degenerate companion,  $\beta=2.5 \times$ specific angular momentum of the orbit \citep{PZ96}. As an additional model variation we set $\beta=1$, in which case the angular momentum loss is equal to specific angular momentum of the orbit. In comparison to the default model, binaries in the $\beta1$-models will loose less angular momentum per solar mass of lost material, and, therefore, their orbits will widen more during non-conservative mass-transfer. As a third model for the angular momentum loss mode, we assume isotropic re-emission (`isore' model). In this case, mass is assumed to leave the system from the position of the mass gainer. As the accretor star is typically more massive than the donor star (to guarantee mass-transfer stability), it caries little angular momentum. So any material that is lost from the binary from the vicinity of the mass gainer caries with it little angular momentum. Thus, typically, this type of angular momentum loss mode leads to even stronger orbital widening than the previous two models. 

Additionally, we pay special attention to the case of accretion onto WDs. We do this for two reasons. Firstly, it is a common occurrence in the formation of DWDs in the considered regime (see upper panel of Fig.\,\ref{fig:channels2}). 
And secondly, the accretion process is more complicated for WDs compared to non-degenerate stars (e.g. MS stars) because of possible thermonuclear runaways in the accreted material on the surface of the WD. The efficiency with which a WD can retain accreted material is highly debated \citep[e.g.][]{Yar05, Bours2013, Wol13, Kat17,Hil20}, which reflects in several orders of magnitude uncertainty in the related supernova type Ia rate \citep{Bours2013, Claeys2014}. The default model of SeBa is based on the retention efficiencies from \cite{Nom07}, \cite{Hac08}, and \cite{Kat99} \citep[model NSKH07 from][]{Bours2013}. Our alternative model adopts the retention efficiencies of \cite{Pri95}, which predict less efficient accretion onto WDs.

\subsubsection{Stellar wind speed}

The last model variation concerns the stellar winds. To model the effect of stellar winds {\sc SeBa} uses different prescriptions for different stellar evolution phases \citep[see][for an overview]{Toonen12}. For low- and intermediate-mass stars they mainly play a role on the red giant branch (RGB) and asymptotic giant branch (AGB). We assume that the companion accretes according to the Bondi-Hoyle model, which depends on the orbital separation, orbital eccentricity and the velocity of the wind material. By default it is assumed that the wind speed is 2.5$\times$ the escape speed, but as an alternative model we adopt the extreme case of the wind speed matching the escape speed. This is appropriate for AGB stars during the late stages of their evolution. 
Note that mass accretion may be more efficient than Bondi-Hoyle at low ratios of the terminal wind velocity to the relative orbital velocity of the system \citep[e.g.][]{Moh07, Moh12, Sal18, Sal19}. Exploring this later possibility is beyond the scope of this paper.

\subsection{Present-day DWD population}

Next, we simulate the temperatures and magnitudes of the local DWD population observable by \textit{Gaia}. This requires assumptions for the DWD spatial distribution and the Galactic star formation history. As the actual observed data are complete roughly up to 200\,pc (cf. Section~\ref{sec:mes}), we can simplify the modeling by considering the thin disk only. We convolve our synthetic DWD models with a constant star formation history setting the age of the thin disk to be of 8\,Gyr, and we distribute binaries homogeneously around the Sun. We set the star formation rate to 3\,M$_\odot\,$yr$^{-1}$ \citep[e.g.][]{Fantin2019}. Note that the star formation rate is a scaling parameters that only affects the total number of DWDs in the simulation.
The spatial distribution is normalized in such way
that a spherical region of radius $r = 200\,$pc centred on the Sun contains a fraction of systems compared to the whole Galaxy equal to $4\pi r^3/3V \sim 0.02$, where $V=5 \times 10^{11}$\,pc$^{-3}$ is the total disk volume.
Finally, using ages -- resulting from the combination of the DWD formation time provided by {\sc SeBa} and constant star formation history -- and distances we compute DWD absolute \textit{Gaia} magnitudes $M_G$ and colors $G_{\rm BP - RP}$ using WD cooling curves of \citet{Holdberg2008, Kowalski2006, Tremblay2011}. We summarise the results for the fraction of DWDs (per single WD) and the absolute number for DWDs with $8<M_G<13$ and $D<170\,$pc, to conform with selection criteria applied in Section~\ref{sec:mes} (cf. Eq.~\ref{eq:hrsel}).

\section{Formation of the gap}

\begin{figure}
  \centering
  \includegraphics[width=0.45\textwidth]{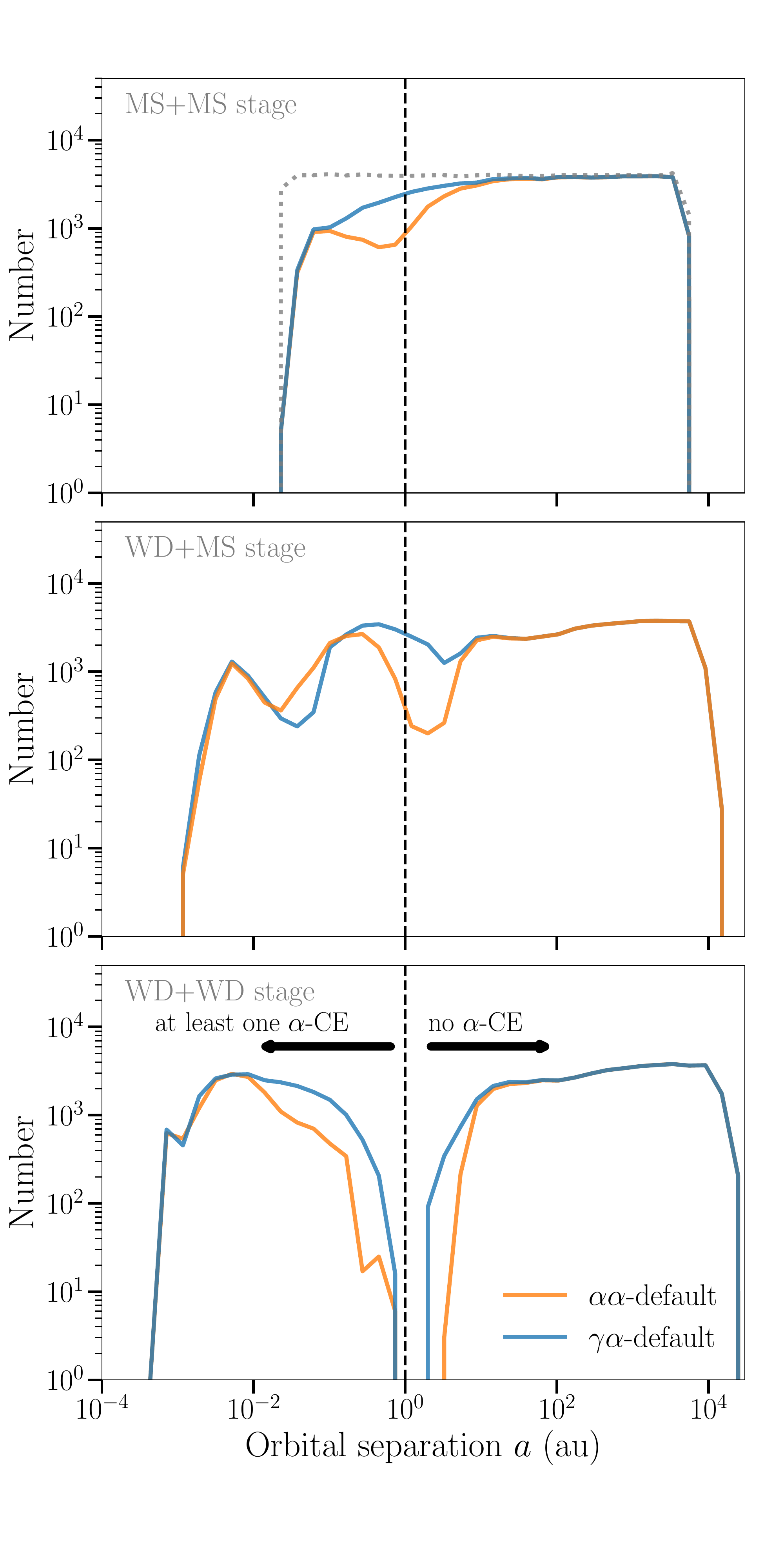}
  \caption[]{Evolution of the orbital separation distribution for two models: $\alpha \alpha$ and  $\gamma \alpha$ represented in orange and blue respectively. From top to bottom we show the initial MS+MS stage, the intermediate WD+MS stage, and the final WD+WD stage. The gray dotted line represent a uniform distribution in $\log (a)$ that has been used to sample MS+MS binaries. It is clearly visible how the distribution expands towards shorter orbital separations from top to bottom. The bottom panel reveals a gap in the distribution at $a \sim 1\,$au. DWDs on the left of the gap have experienced at list one strong orbital shrinkage through the $\alpha$-CE phase. The position of the gap is set by the maximum size that WD progenitor stars can reach late in their evolution.}
   \label{fig:bps_models}
\end{figure}

\begin{figure}
  \centering
  \includegraphics[width=0.45\textwidth]{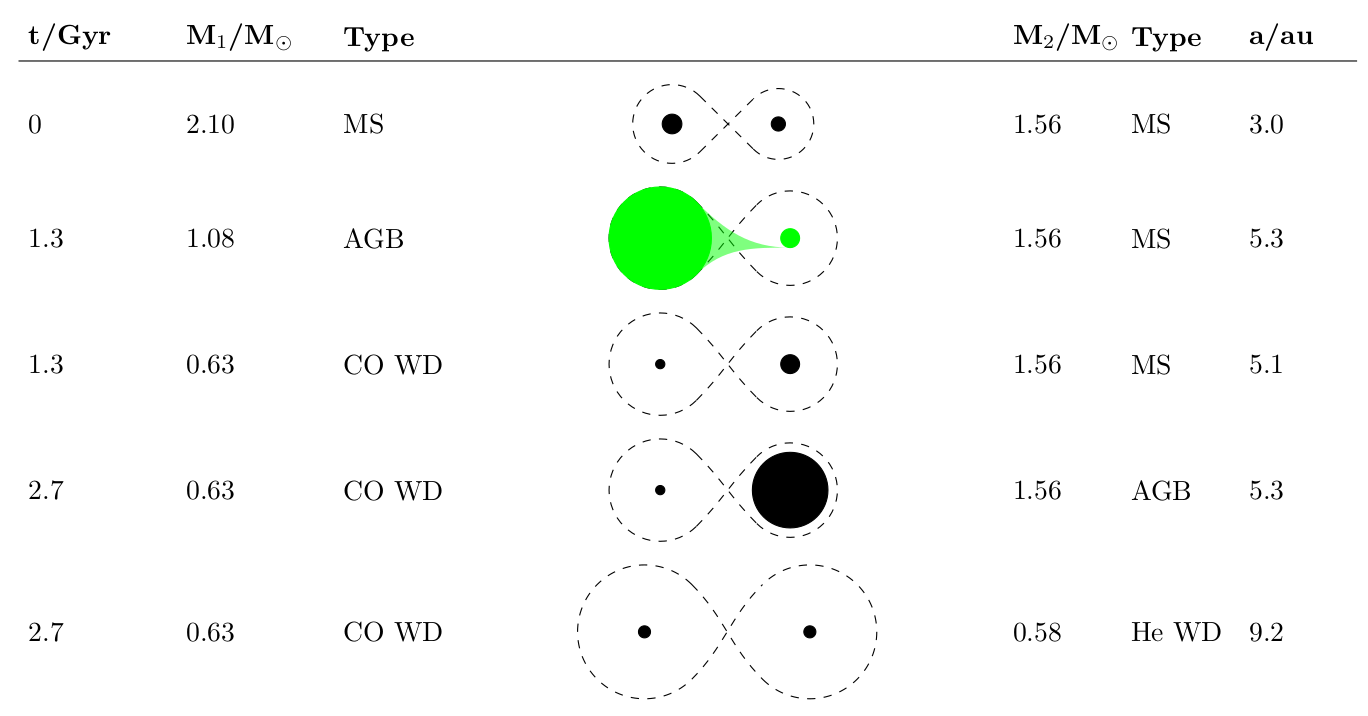}
  \includegraphics[width=0.45\textwidth]{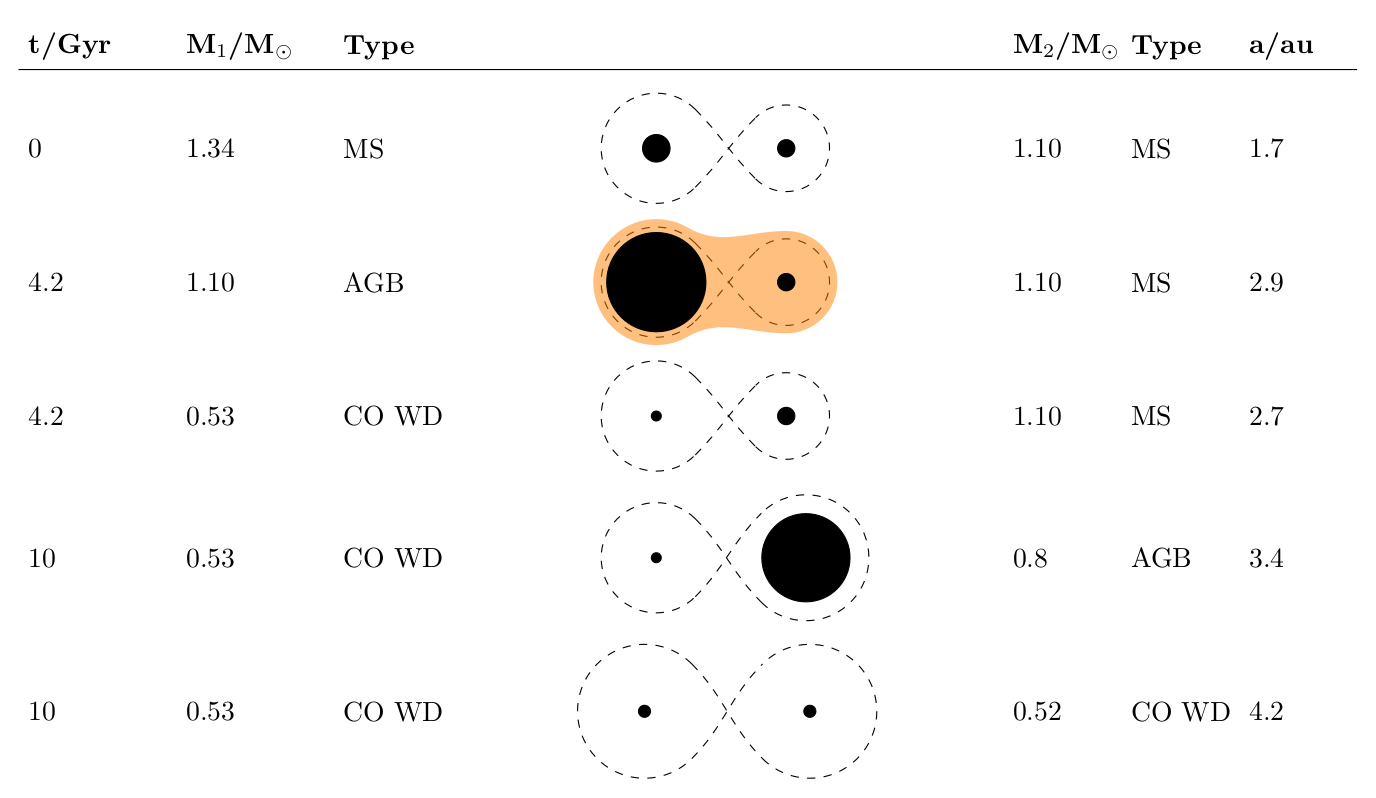}
  \caption[]{DWD evolution channels in which the binary remains above the gap. The top panel is an example of a typical channel in model $\alpha\alpha$, whereas the bottom panel is a typical pathway in model $\alpha\gamma$. Mass-transfer phases are coloured as follows: stable mass-transfer in green, $\gamma$-CE in orange and $\alpha$-CE in gray. Stellar types are abbreviated: main sequence as `MS', red giant branch stars as `RGB', asymptotic giant branch stars as `AGB', Carbon-Oxygen-core WD as `CO WD', Helium-core WD as `He WD'.}
   \label{fig:channels1}
\end{figure}

\begin{figure}
  \centering
       \includegraphics[width=0.45\textwidth]{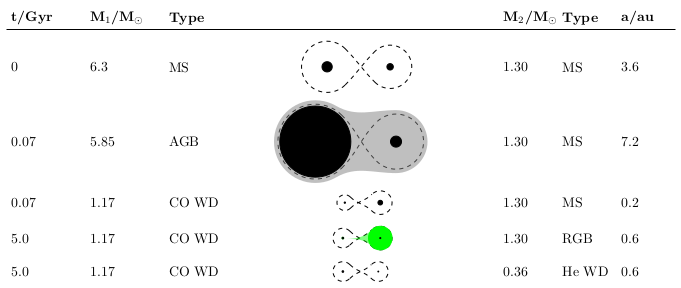}
       \includegraphics[width=0.45\textwidth]{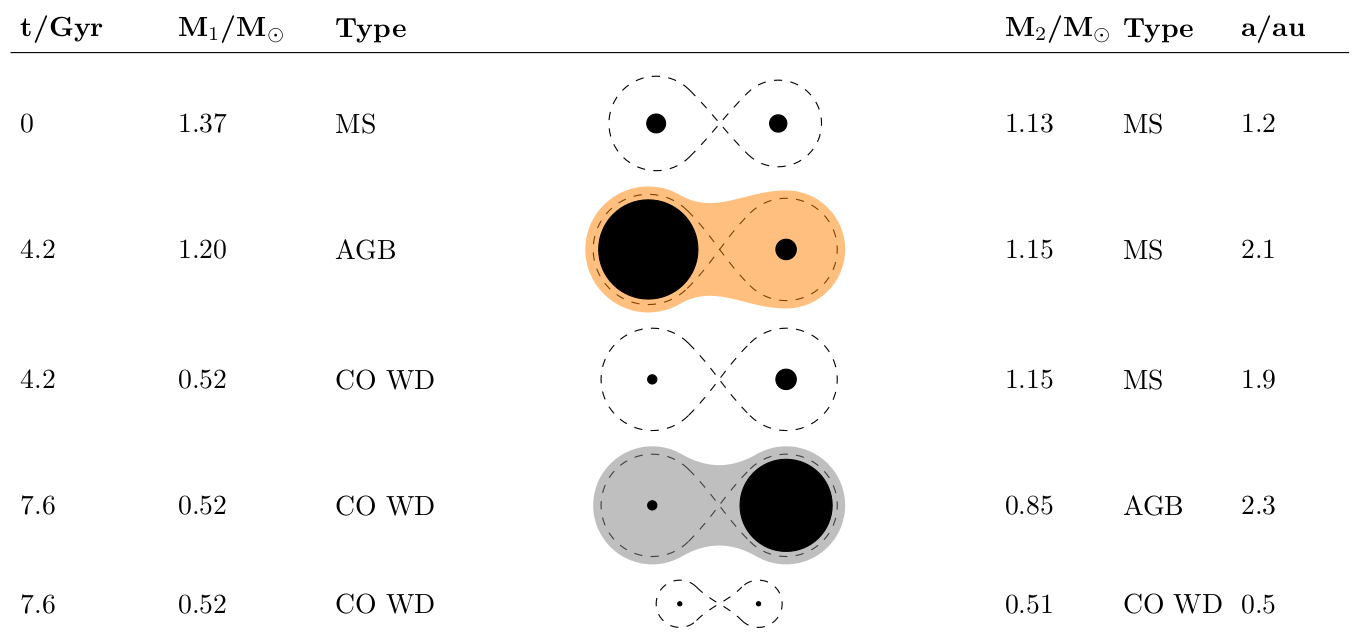}
  \caption[]{DWD evolution channels leading binaries to cross the gap. Colour-coding and abbreviations are the same as in Fig.~\ref{fig:channels1}. The top panel is an example of a typical channel in model $\alpha\alpha$, whereas the bottom panel is a typical pathway in model $\alpha\gamma$.} 
   \label{fig:channels2}
\end{figure}

For those binaries that eventually become DWDs,
Figure~\ref{fig:bps_models} shows the orbital separation distribution at three relevant time steps: initial or MS+MS stage (top panel), intermediate or WD+MS stage (middle panel) when the first WD is formed, and the final or DWD stage (bottom panel) when also the second star becomes a WDs. First, note that, although we draw orbital separations from a flat distribution (gray dotted line), there is a lack of binaries with short orbital separations $< 1\,$au in the top panel. These `missing' binaries have been simulated, but they have merged before the DWD stage, and, thus, are not included in our DWD population. At the WD+MS stage the separation distribution -- initially confined between $10^{-2}$ and $10^4$\,au (top panel) -- expands towards lower values down to $10^{-4}\,$au. In particular, the distribution shows two peaks: one between $10^{-3}$\,au, and another one between $10^{-2}$\,au and at $10^{-1}$\,au. The former arises from a pile up of binaries that undergone a CE phase (cf. Section~\ref{sec:CE}),  while the later from binaries that instead undergone stable mass-transfer (cf. Section~\ref{sec:SMT}). At the DWD stage even more binaries move towards shorter separations after experiencing a CE phase. This leaves a gap at $\sim$1\,au, which is present in both models. Intuitively, one can attribute this migration of binaries towards shorter separation to binary interactions, more specifically, to mass-transfer episodes (stable mass-transfer and/or CE). Binaries with separations larger than several au are too wide for the stars to significantly interact with each other, so they evolve separately (roughly) at the same separation as they were born. On the other hand, binaries with separations of less than a few au experience mass-transfer phases, which change their orbit. The mass-transfer can be initiated when one of the stars evolves off the MS and expands in size such that it becomes larger than the size of its Roche lobe. We find that the division between those binaries that have undergone mass-transfer and those that have not, is  around several au. This scale represents the maximum size that an intermediate-/low-mass star can reach late in its evolution (AGB/RGB stage).

Next, we analyse the  evolutionary pathways that lead a system to end up either above or below the gap more in depth. We find that, besides non-interacting binaries (mainly those with $a(1-e) >> 10\,$au at MS+MS stage), some binaries that undergo one single mass-transfer phase (stable mass-transfer or CE) also end up above the gap at a few au separation. Figure~\ref{fig:channels1} shows their typical evolution for $\alpha \alpha$- (top) and $\gamma \alpha$- (bottom) standard models. These systems start at $a(1-e) \sim 1$\,au. We find that the first -- and the only -- phase of mass-transfer that these binaries experience occurs in a stable manner and is  initiated when the primary fill its Roche lobe late on the AGB experiencing thermal pulses and losing a significant amount of mass through stellar winds. At the end of the stable mass-transfer phase the binary has widened by at most a factor of a few. Although the widening is modest, it prevents the secondary from filling its Roche lobe when it reaches its maximum size on the AGB, and, consequently, from initiating another phase of mass-transfer. We find that even though the final WD remnant mass ranges from 0.4\,M$_\odot$ and 1.4\,M$_\odot$, both the primary and the secondary become WDs immediately without going through a helium stars stage. For the $\gamma \alpha$-model we identify an additional channel leading a binary to separations of a few au at the DWD stage. This channel is very similar to that described above but instead of the stable mass-transfer, the primary initiates a $\gamma$-CE, which allows orbit widening (cf. Section~\ref{sec:CE}). This additional channel represents the reason why $\gamma \alpha$-model presents more binaries at the upper edge of the gap. 

For DWDs that end up at separations $\lesssim 1\,$au the evolution is different with a number of possible channels. However, we find that in all channels across all considered models at least one of the mass-transfer phases is an $\alpha$-CE, which shrinks the binary's orbit severely. Figure~\ref{fig:channels2} shows two examples: one for the $\alpha \alpha$-model (top) and one for the $\gamma \alpha$-model (bottom). The example on the top illustrates the evolution of stars of $6.3\,$M$_\odot$ and $1.3\,$M$_\odot$ on ZAMS, which started $3.6$\,au apart. When the initially more massive star ascends the AGB, the binary's orbit widens by a factor 2 due to the stellar winds. On the AGB the primary fills its Roche lobe and an $\alpha$-CE begins. During the CE evolution the primary loses its hydrogen envelope and becomes a WD of 1.17\,M$_\odot$, and, importantly, the orbit shrinks considerably from 7.2\,au to 0.2\,au. When the initially less massive star evolves off the main sequence, it fills its Roche lobe as a RGB star and initiates a stable mass-transfer. At the outcome of the mass-transfer the secondary becomes a Helium-core WD (He WD) of 0.36\,M$_\odot$, while the orbit of the binary has not changed much. The same channel exists also in the $\gamma \alpha$-model, but its contribution is sub-dominant.
Instead, we find that in the $\gamma \alpha$-model most systems at the lower edge of the gap typically experience two CE phases: a $\gamma$-CE first, followed by an $\alpha$-CE \citep[equivalent to the evolutionary channel derived from observations by][]{Nelemans2000}. The bottom panel of Figure~\ref{fig:channels2} illustrates this channel.
The example system starts as a nearly equal-mass binary with components of 1.3\,M$_\odot$ and 1.13\,M$_\odot$ separated by 1.2\,au. As the primary ascends the AGB, it fills the Roche lobe and initiates the $\gamma$-CE evolution. The system exits the CE phase with the primary being a Carbon/Oxygen-core (CO) WD and a slightly tighter orbit. When the secondary fills its Roche lobe, also on the AGB after losing some of the mass through winds, the system becomes unequal mass, and the $\alpha$-CE develops. Again, after the $\alpha$-CE the orbit decreases severely and a double CO-core WD system is formed. 

To summarise, we find that binaries with initial orbital periods of several au experience at least one phase of mass-transfer. If this is a stable mass-transfer phase or a $\gamma$-CE, the change orbital separation is small (e.g. less than an order of magnitude). However, if the system develops an $\alpha$-CE, it will necessarily shrink and move to significantly (more than an order of magnitude) shorter separations. Therefore, we identify the $\alpha$-CE phase as responsible for opening the gap in the DWD separation distribution. For all considered models and channels therein, we find that stars fill their Roche lobe late in the evolution, either at the AGB or RGB stage. This is a consequence of the size of the orbit and a physical size that a intermediate-/low-mass WD progenitor stars can reach. 

Given our conclusions above, we consider additional model variation to investigate how the efficiency of the CE, the speed of the stellar winds, the efficiency of accretion onto WDs, and the angular momentum loss mode of stable mass-transfer (cf. Section~\ref{sec:DWD_evolution}) affect the location and shape of the gap in the final DWD separation distribution. Importantly, we find that none of the model variations  fill in the gap, which remain present at $\sim 1\,$au in all model variations. The considered variations affect mainly the width and the shape of the gap. In particular, we find that the width of the gap is the largest (0.2 - 3\,au) when increasing the CE efficiency. This is expected, as a higher CE efficiency (lower values of $\alpha \lambda$) shrinks the orbit more severely allowing more binaries to migrate towards shorter orbital periods. As a consequence, more binaries can merge before reaching the DWD stage. Indeed, Table~\ref{tab:models} shows that the DWD fraction for $\gamma \alpha$-2 and $\alpha \alpha$-2 is the lowest. On the contrary, beta1 and isore models reduce the width of the gap to 0.8 - 2\,au. This is because binaries in both models loose less angular momentum, and, consequently, their orbits widen more during non-conservative mass-transfer compared to our default models.

\begin{figure*}
  \centering
  \includegraphics[width=0.45\textwidth]{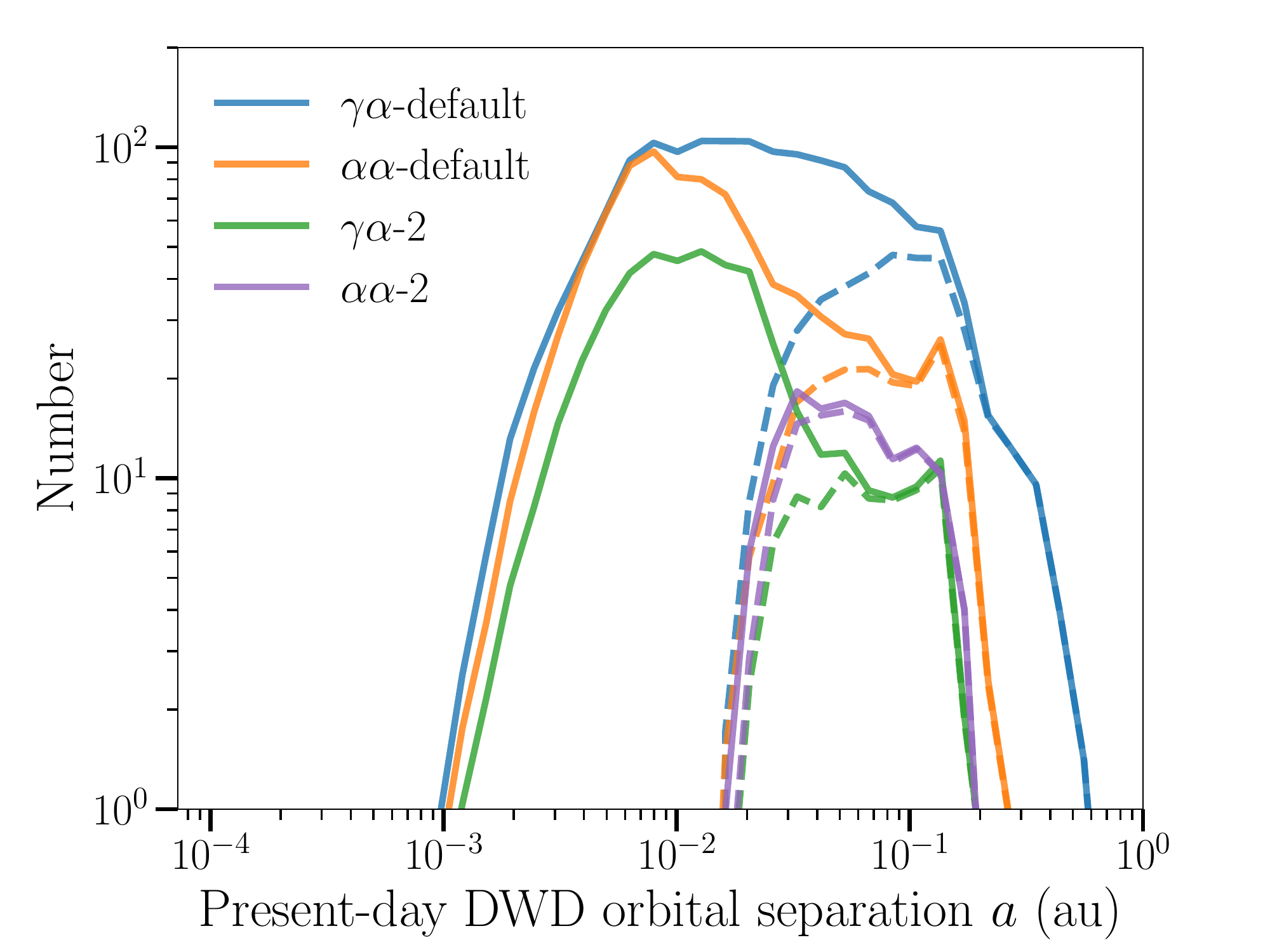}
  \includegraphics[width=0.45\textwidth]{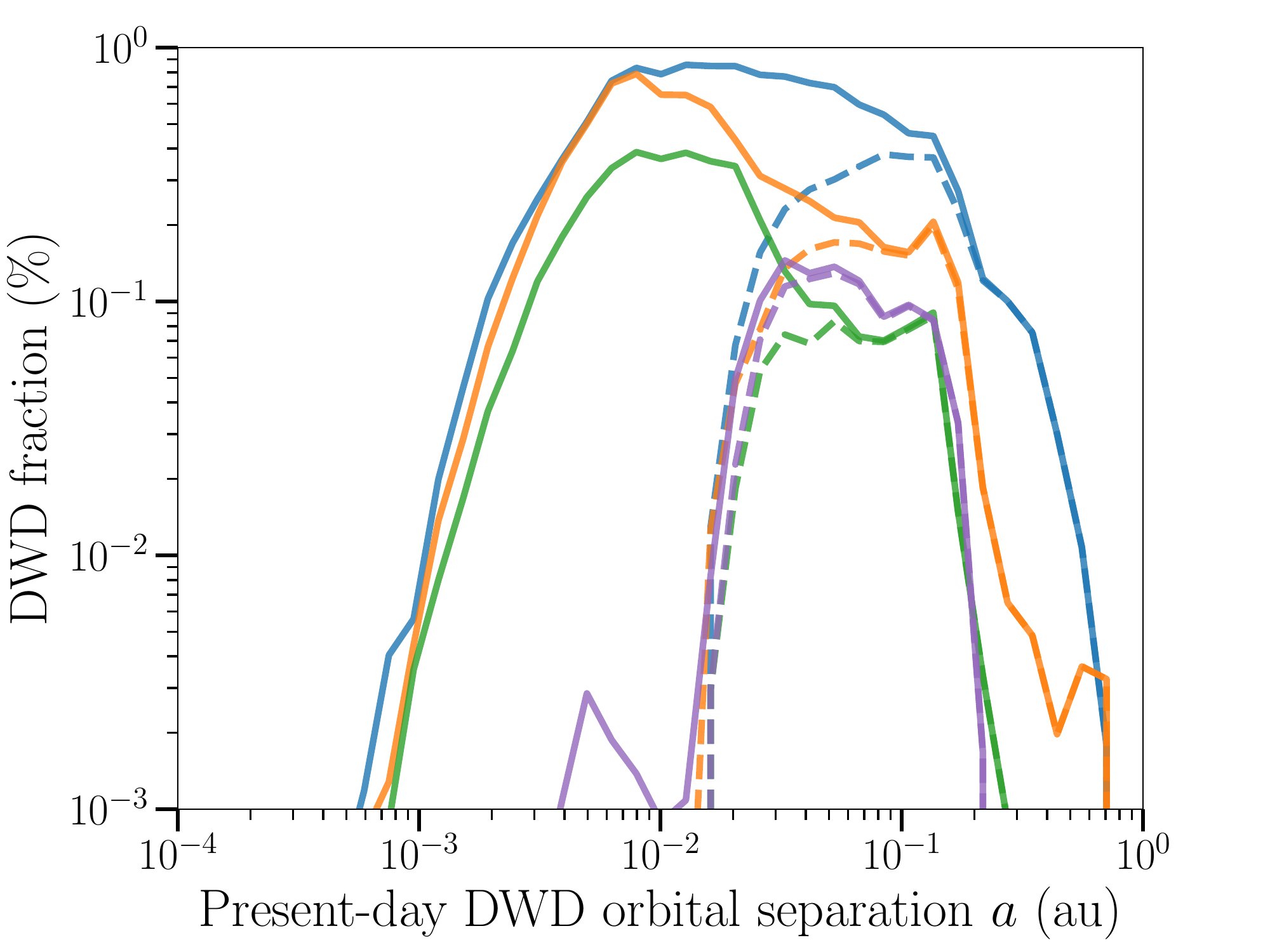}
  \includegraphics[width=0.45\textwidth]{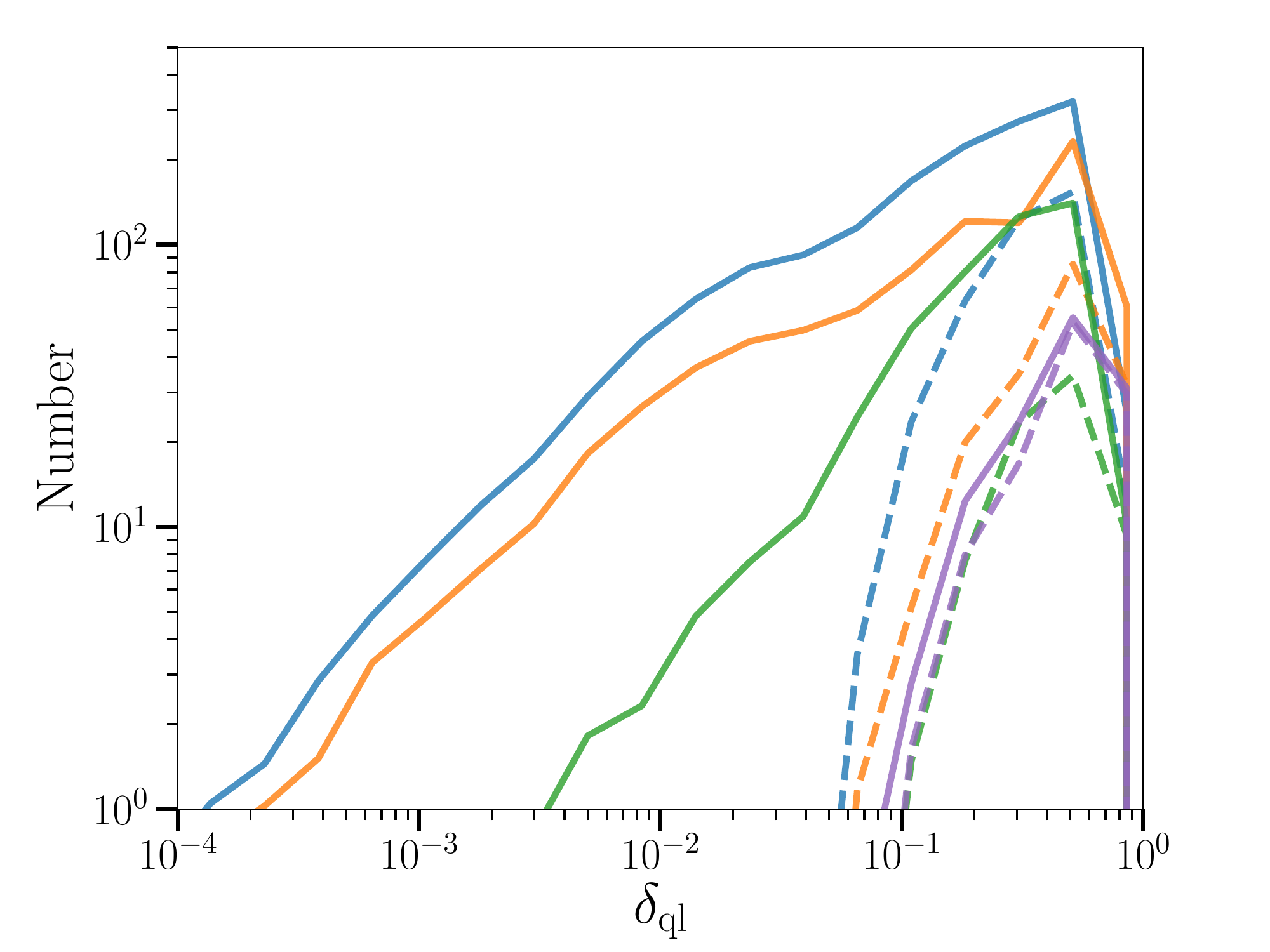}
\includegraphics[width=0.45\textwidth]{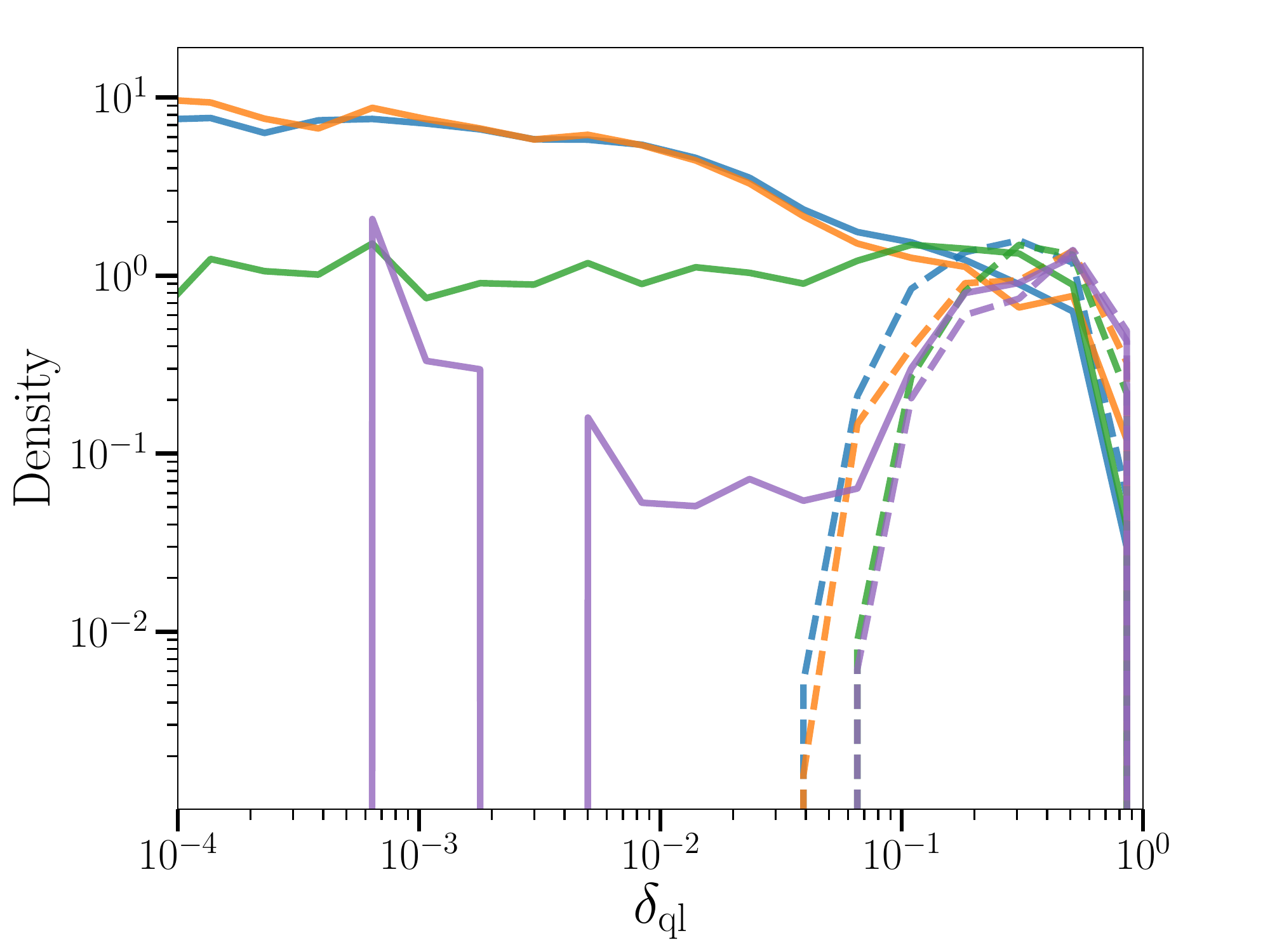}

  \caption[]{Distributions of the present-day DWD orbital separations $a$ in au (top left), DWD fraction as a function of the orbital separation $a$ (top right), and distributions of $\delta_{\rm ql}=|q-l|(q+1)^{-1}(l+1)^{-1}$ (bottom), where $q$ and $l$ are the mass and luminosity ratio correspondingly, in the four models: $\gamma\alpha$- and $\alpha\alpha-$default, and $\gamma\alpha$-2, $\alpha\alpha$-2 with lower CE efficiency.
}
   \label{fig:present_day}
\end{figure*}

\begin{table*}
\centering
\caption{Summary table of synthetic DWD models with $D < 170\,$pc and absolute magnitude $8<M_G<13$. The DWD fraction is the fraction of DWD per single WD in the model. The last two columns list Kolmogorov-Smirnov (KS) and $\chi^2$ statistics resulting from the comparison between \textit{Gaia} data and models.}
\label{tab:models}
\begin{tabular}{lcccccc}
\hline
\multicolumn{1}{|l|}{Model} &
  \multicolumn{1}{c|}{\begin{tabular}[c]{@{}c@{}}DWD fraction \\\end{tabular}} &
  \multicolumn{1}{c|}{\begin{tabular}[c]{@{}c@{}}Number \end{tabular}} &
    \multicolumn{1}{c|}{\begin{tabular}[c]{@{}c@{}}DWD fraction\\ $0.01 < \delta_a < 1$ \end{tabular}} &
  \multicolumn{1}{c|}{\begin{tabular}[c]{@{}c@{}}Number \\ $0.01 < \delta_a < 1$ \end{tabular}} &
  \multicolumn{1}{c|}{KS stat.} &
  \multicolumn{1}{c|}{$\chi^2$ stat.} \\
  \hline
$\alpha \alpha-$default  & 7.0\,\%  & 538 & 2.1\,\% & 108  & 0.43 & 3.02 \\
$\alpha \alpha-2$ & 1.0\,\%  & 76  & 1.3\,\% & 67 & 0.56& 4.08\\ 
$\alpha \alpha-$slow wind & 7.0\,\% & 538 & 2.1\,\% & 231 & 0.5& 3.04\\
$\alpha \alpha-$wd acc. & 12.0\,\% & 911 & 4.4\,\% & 235 &  0.31 & 1.74\\
$\alpha \alpha-\beta1$ & 13.4\,\% & 1016 & 3.8\,\% & 208 & 0.31 & 1.71\\
$\alpha \alpha-$isore & 8.2\,\% & 625 & 2.8\,\%& 147 & 0.37& 2.30\\

& &  & & & &  \\

$\gamma \alpha-$default  & 12.0\,\% & 914 & 4.2\,\%& 233 & 0.31 & 1.74 \\
$\gamma \alpha-2$ & 3.70\,\%  & 283  & 0.8\,\% & 47 & 0.75 & 4.88 \\
$\gamma \alpha-$slow wind & 12.0\,\% & 912 & 4.2\,\%& 231 & 0.31 & 1.73 \\
$\gamma \alpha-$wd acc. & 12.0\,\% & 913 & 4.2\,\% & 235 & 0.31 & 1.74\\
$\gamma \alpha-\beta1$ & 13.4\,\% & 1018 & 3.8\,\% & 208 & 0.31 & 1.70\\ 
$\gamma \alpha-$isore & 12.9\,\% & 985 & 3.6\,\%& 202 & 0.38 & 2.08\\

\hline
\end{tabular}
\end{table*}

\begin{figure*}
  \centering
\includegraphics[width=1\textwidth]{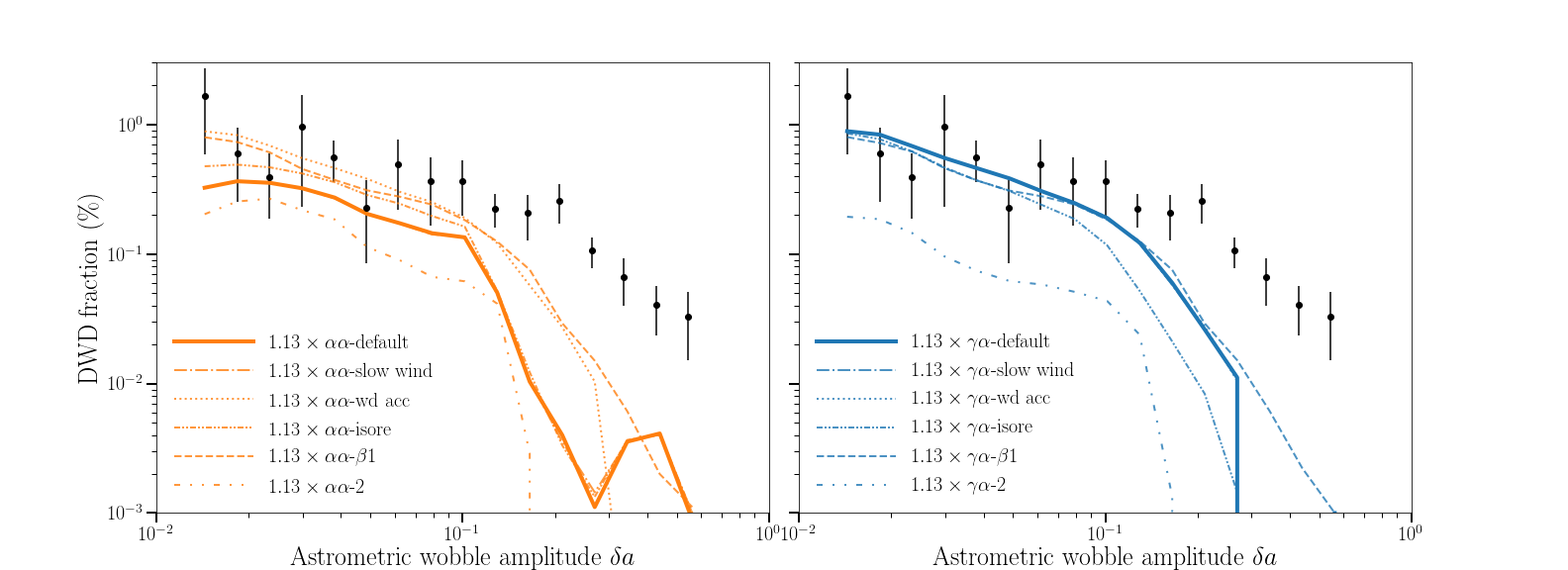}
  \caption[]{Theoretical distributions of the astrometric wobble amplitude constructed base on our two BPS families of models: $\alpha\alpha$ (orange) on the left and $\gamma\alpha$ (blue) on the right. All models are multiplied by a factor of 1.13 that compensates for the difference between observed and simulated (single) WDs. Over-plotted as black circles the observed distribution of the astrometric wobble amplitude.}
   \label{fig:bps_models_wobble}
\end{figure*}

\section{DWD Data vs DWD Models} \label{sec:data_vs_model}

As described in Sections~\ref{sec:select} and~\ref{sec:measure}, multiple selection cuts were applied to the {\it Gaia} EDR3 data to produce a sample of WD systems as free of spurious detections as possible. Most of these cuts result in a straightforward reduction of the total number of DWD binaries leaving the shape of the distribution of the astrometric wobble amplitude $\delta a$ unchanged. Because the same set of selection criteria is applied to the sample of all WDs (single stars or those with companions), the relative fraction of DWDs recovered remains unaffected (as demonstrated for example in Figure~\ref{fig:dwd_da}). Nonetheless, as pointed out in Section~\ref{sec:wa_dist}, {\it Gaia} EDR3's sensitivity to the photometric centroid shift drops quickly as the amplitude of the wobble shrinks to match the uncertainty of the astrometric measurement (a quantity dependent on various observational parameters, e.g. the apparent magnitude of the star). Similarly, as the period of the binary grows beyond the time span of the {\it Gaia} EDR3 observations, the sky projection of the portion of the orbit tends to appear more linear and is therefore absorbed into the proper motion value reducing (or removing) the excess in the goodness-of-fit statistic. Above we argued that the transfer of power from the RUWE-based statistics to the proper motion anomaly happens mostly for systems with separations exceeding 2\,au, i.e. in the regime not studied here. However, the drop of sensitivity at low wobble amplitudes needs to be taken into account. As we describe below, this is easy to implement with the simulated DWDs in hand. In practice, we assume that {\it Gaia}'s DWD detection efficiency is a step function with no systems detected below $\delta a=0.01$.

Figure~\ref{fig:present_day} exhibits distributions of the present-day DWD orbital separations $a$  in au (top left), DWD fraction as a function of the orbital separation (top right) as well as distributions of the $\delta_{\rm ql}$ values (bottom) in the four models (default and those with lower CE efficiency). Additionally, we select DWD systems with heliocentric distance $D < 170\,$pc and absolute magnitude $8<M_G<13$ (solid lines). Note that in the Figure, dashed lines show the same distributions modified by taking into account the wobble amplitude threshold of $\delta a> 0.01$. As expected, only systems with relatively large separations and/or large values of $\delta_{\rm ql}$ can currently be picked up by {\it Gaia}. In three out of the four displayed models (all but $\alpha\alpha-2$), this selection effect leaves the bulk of the DWD systems, i.e. those that have shrunk below $a=0.01$\,au, undetected. Note that while the undetected DWD systems have smaller separations compared to the detected ones (by factor of 5 to 10), the main difference is in their $\delta_{\rm ql}$ values. The undetected binaries typically have between 10 to 100 times smaller $\delta_{\rm ql}$. Importantly, the distributions of the $\delta_{\rm ql}$ factor are very similar across the four models: all distributions peak around $\delta_{\rm ql}\approx0.5$, show a sharp truncation around $\delta_{\rm ql}\approx 1$ and very long tail towards tiny $\delta_{\rm ql}<10^{-3}$ (see bottom panel of the Figure where density is shown per bin of $\delta_{\rm ql}$).

In Table~\ref{tab:models} we report the DWD fraction and the total number of DWDs with $D < 170\,$pc and absolute magnitude $8<M_G<13$, and for the sub-sample with $0.01 <\delta_a < 1$ in our synthetic models. Considering the whole simulated orbital separation range, the DWD fraction varies between 1\,per cent and 13.4\,per cent across models. The family of $\gamma \alpha$ models presents higher DWD fraction, except for `$\beta$-1' and `wd. acc' model variations for which the fraction is the same for both families of model. In particular, $\alpha\alpha$-2 and $\gamma\alpha$-2 models, in which the binary orbit shrinks strongly during the CE evolution, yield the lowest DWD fractions. When restring to the sub-sample with $0.01 <\delta_a < 1$, the DWD fractions decrease. Based on the {\it Gaia} sample we measure the overall fraction of DWDs of $(6.5 \pm 3.7)\,$per cent in the regime of $0.01<\delta a<1$ (corresponding to orbital separations $0.01 < a\,(\text{au}) < 2$, cf. Section~\ref{sec:wa_dist}).
Note, however, that the $\alpha\alpha$-2 model that shows a small increase in the DWD fraction. In the $\alpha\alpha$-2 model, binaries with short separations have already merged, while most of the binaries survived until the present day are within the range accessible with {\it Gaia} through astromentric wobble. We find that all models predict lower DWD fractions than 6.5\,per cent derived from the data. This is because the total number of DWDs with $0.01 <\delta_a < 1$ (in our synthetic models) is at least twice as low compared to the total DWD number deduced from the data (555), while the number of single WDs (simulated vs observed) is more similar (7605 vs 8568). We can artificially adjust the predicted DWD numbers thought a factor $8568/7605 = 1.13$ accounting for the difference between observed and simulated single WDs.

Figure~\ref{fig:bps_models_wobble} compares the observed distribution of the astrometric wobble amplitude $\delta a$ to that of the theoretical prediction of our two families of models: $\alpha\alpha$ on the left and $\gamma\alpha$ on the right.  For both panels, the normalisation of synthetic models (resulting from the combination of the star-formation, the initial binary fraction, as well as the selection cuts applied) is adjusted by the 1.13 factor defined above. All models predict the existence of a deep gap, albeit with somewhat different profile shape. Comparing Figure~\ref{fig:bps_models_wobble} to Figures~\ref{fig:present_day} and ~\ref{fig:bps_models}, it is clear that the sharp fall in the fraction of DWD systems beyond $\delta a \approx 0.2$ corresponds to the left side of the gap carved out in the distribution of DWD separations by the binary interactions. The observed distribution of $\delta a$ values thus probes the pile-up of the systems that have shrunk substantially and are observed today at much lower values of $\delta a$ (and $a$). To our knowledge this is the first detection of the break in the DWD separation distribution corresponding to a gap produced (mostly) by the CE evolution.  Note that due to the limitations of the method, currently only one side (at low $\delta a$ and $a$) of the gap can be mapped. The models predict that the fraction of DWDs starts to recovers beyond a few au (see Figure~\ref{fig:present_day}) and for systems wider than $a=10$\,au appears to follow the initial distribution of the DMS separations \citep[][]{Elbadry2018,Torres2022}.

All models but $\alpha \alpha$-2 and $\gamma \alpha$-2 follow the data up to $da=0.1$. At larger $\delta a$ the data show a shallower decline compared to the models. In Table~\ref{tab:models} we report results of the Kolmogorov-Smirnov (KS) and $\chi^2$ tests. Base on the KS-test only $\alpha \alpha$-2, $\gamma \alpha$-2 and $\alpha \alpha$-slow wind can be considered significantly different from the data (at 0.05 significance level). While the $\chi^2$-test reveals the preference for the $\gamma \alpha$ set of models, for which the $\chi^2$ statistic is lower compared to the same model variations in the $\alpha \alpha$-family (except for `wd. acc' and `$\beta1$' variations). Models $\alpha\alpha$-wd acc, $\alpha\alpha-\beta$1, $\gamma\alpha$-default, $\gamma\alpha$-slow. wind, $\gamma\alpha$-wd acc, $\gamma\alpha-\beta1$ show the lowest $\chi^2$ values. Figure~\ref{fig:bps_models_wobble} illustrates that these models better reproduce the data for $\delta a < 0.1$, and display a long smooth decrease in DWD fraction for $\delta a > 0.1$.

\section{Discussion and Conclusions} \label{sec:discussion}

Using excess in the goodness-of-fit of the astrometric solution in the {\it Gaia} EDR3 data we measure the amplitude of the angular wobble for hundreds of nearby unresolved double white dwarf systems (DWDs). Multiplied by (parallax-based) distance, the angular wobble amplitude can be translated into physical units, $\delta a$ - a quantity proportional to the binary separation in au. Relying on the `strong wobble' systems only, i.e. those in the regime unaffected by random fluctuations of the reduced $\chi^2$, we build the distribution of the wobble amplitude $\delta a$. We detect a sharp drop in the numbers of DWD systems with $\delta a > 0.2$ ($\log_{10}\delta a>-0.7$). The break in the $\delta a$ distribution appears robust against the WD selection criteria used. 

Relating $\delta a$ values to the corresponding separations $a$ requires an assumption about the mass-luminosity behaviour of the DWD components. Using synthetic DWD populations we demonstrate that the break in the $\delta a$ distribution maps onto a break in the distribution of binary separations, $a$. In fact, the break discovered is one half (the left-hand side) of a gap predicted in the distribution of DWD separations (cf. Figure~\ref{fig:bps_models}, see also \citealt{Toonen14}). In the considered models, a prominent gap is produced by systems starting at separations of several au and experiencing strong interactions during the evolution. Most of the systems that undergo CE evolution shrink their sizes and migrate to separations $\ll$1\,au, thus clearing out a gap. Note that the current {\it Gaia}-based measurements of the astrometric wobble lack sensitivity in both low ($\log \delta a<-2$) and high ($\log\delta a >0$) amplitude regime. Interpreting the observational constraints with the help of theoretical models, we predict that the bulk of the small-separation DWD systems that have undergone interactions are hiding at $a<10^{-2}$ au, currently below the astrometric detectability threshold. For separations greater than tens of au, the stellar components of a DWD evolve independently from each other, therefore we predict that beyond the gap the number of DWD binaries will quickly grow in accordance with the period distribution of their progenitor systems \citep[see e.g.][]{Raghavan2010, Badenes2018} and have been confirmed by observations of DWDs in astrometric pairs \citep[e.g.][]{Elbadry2018,Torres2022}.

By integrating the wobble amplitude distribution $\delta a$, we estimated the DWD fraction of $(6.5 \pm 3.7)\,$per cent, which is in agreement with other measurements in the literature as we detailed in the following.
So far, most measurements of the DWD fraction have been derived from spectroscopic samples. One of the first estimated by \citet{Maxted_1999} reports the DWD fraction of $1.7-19$~per cent (with 95~per cent confidence), which is based on a sample of 46 WDs. This result became more precise as the sample of known (single) WDs and DWDs increased with time. It is important to highlight that all measurements of DWD fraction are limited to a separation range defined by the selection effects of the method. Based on the combined SDSS and SPY samples, \citet{Maoz2018} derived a fraction of $(9.5 \pm 2)$\, per sent for DWD orbital separations up to 4\,au. More recently, \citet{Napiwotzki2020} estimated a binary fraction of $(6 \pm 1)$\,per cent based a larger sub-sample of 625 WDs from SPY; note that the SDSS sample is sensitive to DWD separations of at most 0.1\,au. When the result of \citet{Maoz2018} is re-scaled to the same separation range, the two measurements are in agreement; when re-scaled to $0.01 < a\text{(au)} < 1$ \citep[using Equation~8 of][]{kor21} their DWD fraction decrees to $\sim 7.5\,$per cent. Note that the analysis of \citet{Maoz2018} assumes a continuous power law for the DWD separation distribution, however, the presence of the gap in the separation distribution would imply that their DWD fraction could be higher than originally reported. 

We compared the shape of the observed distribution of the astrometric wobble $\delta a$ to that resulting from BPS models (cf. Section~\ref{sec:data_vs_model}). The comparison revealed that our models qualitatively agree well with the data up to $\delta a \simeq 0.1$, but at $\delta a > 0.1$ the number of simulated DWD precipitates more quickly than the data (e.g. Figure~\ref{fig:bps_models_wobble}). We found that by multiplying the synthetic $\delta a$ by a factor of 2 - 3 we could bring most of models in agreement with the data (cf. bottom panels of Figure~\ref{fig:bps_models_wobble_fit}). Physically, this suggests that orbital separations in our DWD evolution models should shrink less. 

\subsection{Possible reasons for the discrepancy}

There can be various plausible reasons for the discrepancies  between the observed astrometric wobble amplitude distribution and that predicted by our synthetic models. The apparent mismatch may be related to the observational biases in our DWD sample. Alternatively, the discrepancies could already be hidden within the adopted initial conditions or occur subsequently in the parametrisation of processes involved in the binary evolution. We discuss some possibilities below.

\textit{Observational biases and sample contamination.} Using the results of \citet{Penoyre2021}, \cite[see also][]{Penoyre2020, Penoyre2022}, we estimate the selection efficiency of the astrometric wobble method not to vary dramatically in the range of $\delta a$ considered here. Note however that as shown in Figure 15 of  \citet{Penoyre2021} the drop in efficiency around $\log \delta a\approx -2$ is fast and thus may have affected the lowest amplitude portion of the wobble distribution more than envisaged. On the other hand, noticeable but a more gentle decrease in the fraction of predicted detections is seen for systems with $\log \delta a>-1$. This would imply that the slope of the distribution beyond the break is  shallower and the mismatch between the data and the models in this regime is even worse. Systems with more than 2 companions could contribute to lift up the $\log \delta a>-1$ wing of the distribution. As discussed in \citet{B20} and \citet{Penoyre2022}, triples and higher multiples tend to provide objects with the highest detected wobble amplitude $\delta a$. Note that to date, very few observational constraints exist for WDs in triples \citep[see e.g.][]{Toonen2017} or for triple WD systems \citep[see e.g.][]{Perpi2019}. Discovering such systems with {\it Gaia} would be truly exciting. Finally, systems with massive dark companions, i.e. WD+NS or WD+BH would easily masquerade themselves as DWDs but would exhibit a more pronounced wobble. Other exotic systems such as two WDs co-orbiting with a NS similar to that discovered by \citet{Ransom2014} can not be ruled out either.

\textit{Common envelope efficiency.} A straightforward conjecture would be that the CE efficiency (i.e. the $\alpha$ parameter) is higher than assumed here, leading to a less severe shrinkage of the orbit for the sources just below the gap.
However, one should be careful with such a claim as a reduced efficiency of the CE phase, can severely change the evolutionary channels. For instance, for the example system of model $\alpha\alpha$ in the top panel of Figure\,\ref{fig:channels2}, if the post-CE orbital separation is much wider, the secondary star would be more evolved when it initiates the second phase of mass-transfer, such that it would likely not occur in a stable manner (as in the figure), but lead to a second CE phase. In this case the same binary would end up as a much more compact DWD even though the CE occur more efficiently.  For the example system of model $\gamma\alpha$ (bottom panel of Figure\,\ref{fig:channels2}), an increased CE efficiency would instead give rise to wider DWD orbits. 

\textit{Binding energy of the envelope.} Instead of an increased CE efficiency, the same effect can be achieved with a reduced the binding energy of the envelope, i.e. an increase in $\lambda$. In our default model we have assumed a relatively high constant value of $\alpha \times \lambda =2$, whereas the classical assumption in binary population synthesis is $\alpha \times \lambda =1$. High(er) values of $\lambda$ are reasonable for the progenitors of the DWDs just below the gap \citep[e.g.][]{van10}; these systems are formed by mass-transfer phases with donor stars in the late stages of evolution on the AGB, when their stellar envelopes can be weakly bound to the stars. If so, the value of $\lambda$ may be larger than we assume here, which in turn would help to get larger orbital separations.

\textit{Stable mass-transfer.} Besides the modelling of the CE phase, we have constructed models in which other physical processes are varied (cf. Section~\ref{sec:bps}). The model variations that match best with the observations presented here are the `$\beta1$'-variation and the `wd acc'-variation which are both related to modelling stable mass-transfer (cf. Section~\ref{sec:SMT}).
The former (`$\beta1$') model variation concerns the angular momentum loss mode, i.e the parametrisation of how much angular momentum is lost when the mass-transfer is not conservative. Unfortunately, this assumption has been poorly constrained \citep[see e.g.][]{Toonen14}, and so it is interesting to add that astrometric wobble data favour small levels of angular momentum loss in order to widen the orbits. The latter (`wd acc') variation concerns the efficiency of mass accretion onto a WD during mass-transfer. The model preferred by the {\it Gaia} data suggests that WDs could accrete less efficiently than in the default model. This has important consequences, most notably for supernova type Ia progenitors: if it is  harder for a WD to grow in mass, fewer WDs will reach the Chandrasekhar mass due to mass-transfer \citep[see e.g. ][]{Bours2013}.

\textit{Initial orbital period distribution.} We also tested how the choice of the initial orbital separation distribution influences the size and the shape of the gap. We constructed an additional model variation (for both $\alpha\alpha$ and $\gamma\alpha$ families) in which we draw orbital periods from a log-normal distribution with a mean at 5.03\,days, a dispersion of 2.28\,days \citep{Raghavan2010}, and up to a maximum period of $10^{10}$\,days. Note that for a system with the total mass of 1.5\,M$_\odot$, the distribution peaks at 1--2\,au, i.e exactly where we expect the gap to be formed. We found that at the DWD stage the distribution of orbital separations differ mainly at $a \lesssim 0.1$\,au and $a >$ several au, while the position and the shape of the gap -- mainly carved by the $\alpha$-CE -- to be the same as in the default model. Thus, this model variation does not produce significant differences in the regime accessible through the astrometic wobble in the {\it Gaia} data.

\textit{Eccentricity.} 
In our synthetic models binaries circularise during the CE evolution, however, as this is one of the lest understood phases in binary evolution, in reality this may not necessarily be the case for all binaries. Interestingly, many wide-orbit (100-$10^4$ d) post-mass-transfer systems, show eccentricities ranging from 0.1 to about 0.6 \citep{Jor98, Mat09, Jor16,Han16,Vos17, Esc20}. It is not understood where the eccentricity comes from - tides just before and during the mass-transfer should have circularised the system \citep[but see][]{Bon08,Izz10,Der13,Raf16,Oom20}. 

As demonstrated in \citet{Penoyre2020} and \citet{Penoyre2021} increasing binary's eccentricity works to reduce the amount of astrometric wobble. This effect is illustrated in Figure 3 of \citet{Penoyre2021} and is summarised in their Equation 12. Two factors combine to modify the amplitude of the astrometric perturbation with varying eccentricity. An eccentric orbit has a preferred direction, i.e. the 3D orientation of its major axis. As a result, changing the viewing angles of the binary changes the projection of its orbit on the sky and can alter the amplitude of the centroid excursion. Additionally, unlike for circular orbits, motion along the ellipse is non-linear in time. Thus for binary periods close to the mission baseline, the amplitude of the binary signal will depend on which orbital phase gets sampled. Given these effects, for two binary samples with identical binary separation distributions, a sample of systems with non-zero eccentricity would produce lower on average $\delta a$ values and, consequently, would be interpreted by us to have shorter intrinsic separations $a$. This of course would only make the model-data discrepancy worse. When analysing the data, we assumed that most of DWDs have low eccentricity; this assumption however must be tested. In principle, given that proper motion anomaly and $\chi^2$-excess show different dependence on eccentricity, in the future, DWD eccentricity may be constrained for systems with both high RUWE and PMA.

{\it Gaia}'s efficiency of detecting a wobble in the centre of light of an unresolved binary system will improve with time for both small and large separation systems. Currently, short-period binaries fall below the sensitivity level dictated by the astrometric error and the number of visits per source, i.e. the number of {\it Gaia} measurements. While single-epoch centroiding error is not going to evolve dramatically over time, the number of visits will keep accumulating thus helping to detect - with high significance - small excesses in the goodness-of-fit statistic. Thus there is real hope to use future {\it Gaia} releases to tap into the pile-up of short-separation systems with $a<10^{-2}$\,au. For large separations, astrometric deviations induced by binary motion are typically quasi-linear and can be described by the proper motion component of the astrometric model, therefore leading to an improved single-source model fit for a non-single source, i.e. a reduction in the $\chi^2$ excess and correspondingly in $\delta a$. As the temporal baseline of the mission grows, larger non-linear portions of the long-period binary motions are sampled resulting in the transfer of power from the proper motion anomaly to the RUWE-based statistics. Little is known about systems on the other side of the gap with $1<a($au$)<10$ as other techniques lack sensitivity in this regime: RV-based binary detection requires shorter orbital periods \citep[see e.g.][]{Maoz2018, Napiwotzki2020}, while typical common proper motion pairs with {\it Gaia} are presently limited to $a>100$ au \citep[][]{Elbadry2018, Elbadry2021}. In the future, we hope to see the DWD systems tracing the right-hand side of the gap in the distribution of separations. Our work shows that the {\it Gaia} data has the potential to fully map the gap in the DWD separation distribution providing an important benchmark for testing binary evolution models and our understanding of the physical processes involved in shaping binaries' orbits.

\section*{Acknowledgments}

VK and ST acknowledges support from the Netherlands Research Council NWO (Rubicon 019.183EN.015, VENI 639.041.645, VIDI 203.061 grants).

This research made use of data from the European Space Agency mission Gaia
(http://www.cosmos.esa.int/gaia), processed by the Gaia Data
Processing and Analysis Consortium (DPAC,
http://www.cosmos.esa.int/web/gaia/dpac/consortium). Funding for the
DPAC has been provided by national institutions, in particular the
institutions participating in the Gaia Multilateral Agreement. This
paper made used of the Whole Sky Database (wsdb) created by Sergey
Koposov and maintained at the Institute of Astronomy, Cambridge with
financial support from the Science \& Technology Facilities Council
(STFC) and the European Research Council (ERC). 

\section*{Data Availability}

This study is based on published results. A catalogue of DWD candidates analysed in this study will be provided upon acceptance of the manuscript.

\bibliographystyle{mnras}
\bibliography{references}

\label{lastpage}

\appendix
\section{Deconvolving the measured double MS and WD systems}
\label{ref:dwd_dms_detail}

\begin{figure*}
  \centering
  \includegraphics[width=0.99\textwidth]{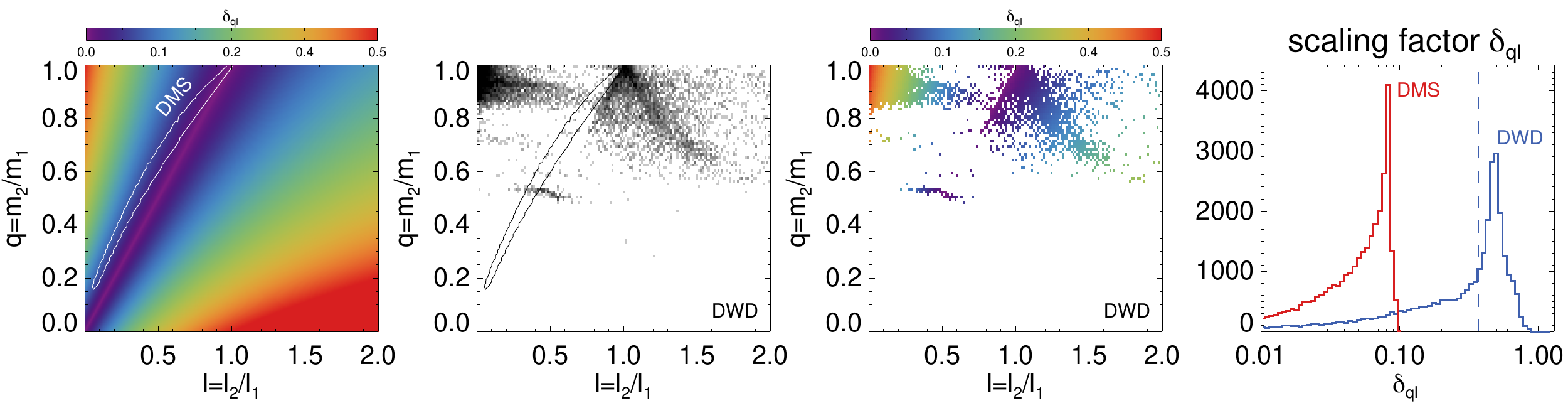}
  \caption[]{{\it 1st panel:} Distribution of the astrometric wobble
    scaling factor $\delta_{al}$ in the space of the luminosity ratio
    $l$ and mass ratio $a$. White contour marks the locus of double MS
    systems with metallicity [Fe/H]=-0.25 and age $10^{1.5}$ Myr. {\it
      2nd panel:} Distribution of DWD systems in the plane of
    $\{q,l\}$. Note the difference in the behaviour between DMSs and
    DWDs. {\it 3rd panel:} Scaling factor $\delta_{ql}$ for DWDs in
    plane of $\{q,l\}$. {\it 4th panel:} Distribution of $\delta_{ql}$
    for DMS (red) and DWD (blue) systems. Typically, $\delta_{ql}$ for
    DWD is some 5 times larger than that of DMS.}
   \label{fig:dql}
\end{figure*}

There is one important difference between the WDs and the MS stars
that ought to be considered before the astrometric properties of the
binary systems are analysed. Rather than being in a state of the
hydrostatic equilibrium between the thermal gas pressure and gravity,
the pull of gravity inside a WD is balanced by the electron degeneracy
pressure. Thus, the WD structural properties differ from that of any
normal star. For example, the more massive WDs are fainter rather than
brighter. This ``inverted'' mass-radius relation is of particular
relevance when calculating the astrometric wobble scaling factor
$\delta_{ql}$. To this end, Figure~\ref{fig:dql} presents the
behaviour of DWD and DMS systems in the plane of luminosity $l$ and
mass $q$ ratios. Similar to Figure 3 of \citetalias{B20}, the left
panel shows the distribution of $\delta_{ql}$ as a function of $q$ and
$l$. Note, however, that we consider the luminosity range shown
to $l=2$. As discussed in \citetalias{B20}, the MS stars obey a simple
powe-law-like luminosity-mass relation. This is demonstrated here with
a white contour showing the locus of model DMS systems created using
random samples from a PARSEC \citep[][]{parsec} isochrone with
[Fe/H]$=-0.25$ and age of $10^{1.5}$ Myr. Note that we have limited
the companion masses to $<0.6$\,M$_{\odot}$ given the absolute magnitude
cut imposed above. As the next (2nd) panel of Figure~\ref{fig:dql}
shows, the evolution of DWD systems in the mass ratio-luminosity ratio
plane is nearly perpendicular to that of the DMSs. This means that
DWDs sample regions of $\{q,l\}$ space with significantly higher
$\delta_{ql}$ values as confirmed in the 3rd panel of the Figure. The
rightmost (4th) panel of Figure~\ref{fig:dql} shows the resulting
distributions of model DWD and DMS systems. We use synthetic  $\gamma\alpha$ DWD models described in Section~\ref{sec:bps}.  Both distributions have
long tails stretching to very low values of the scaling factor,
i.e. $\delta_{ql}<10^{-2}$. However, the peak of the $\delta_{ql}$
distribution for the DWDs is some 5 times higher compared to that of
the DMSs. This is the consequence of the WD mass-luminosity relation
and implies that at fixed $\delta a$, the corresponding DWD
separations are typically much smaller compared to DMS binaries.

\section{Considerations on the mismatch between data and models}

Ideally, to find a model that closely follows the data would require generating a large set of synthetic models by fine-tuning various `knobs' in the binary population synthesis procedure, many of which are correlated and/or degenerate (examples discussed in Section~\ref{sec:discussion}). Thus, this is a high-dimensional and computationally extensive problem, and is out of the scope of the current paper. Here we consider in what direction synthetic models should change to overlap with the data. 

\begin{figure*}
  \centering
\includegraphics[width=1\textwidth]{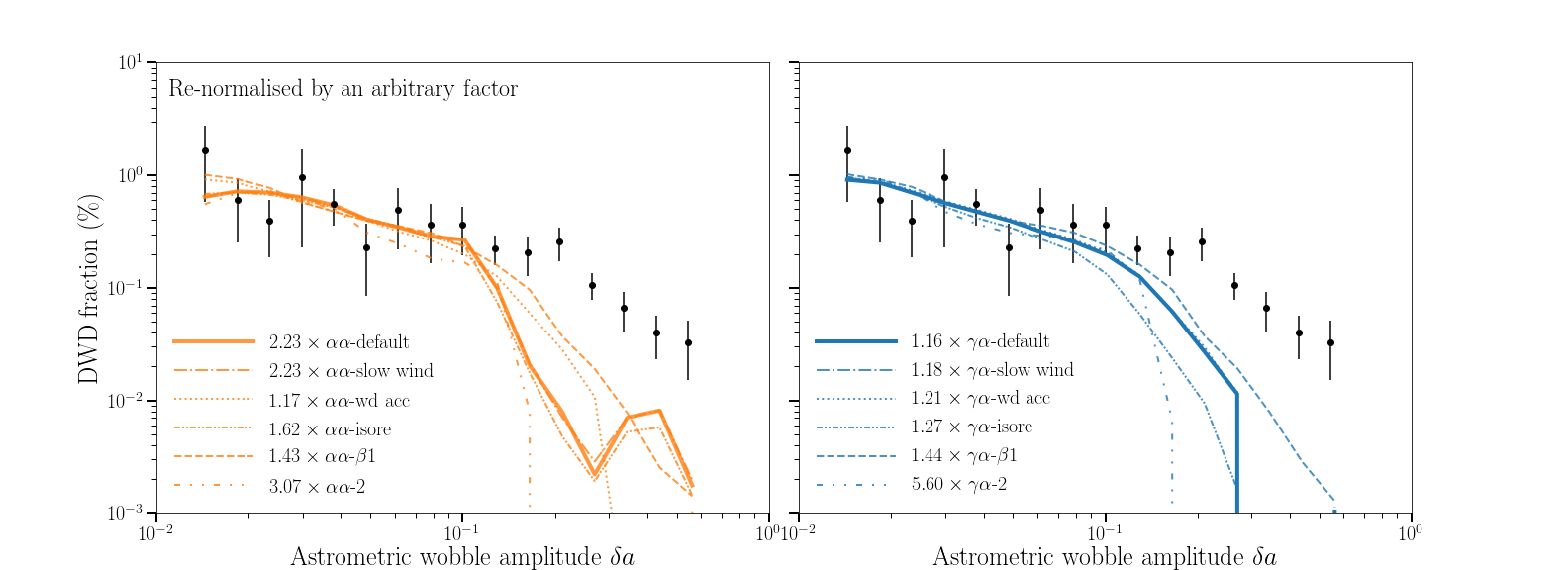}
\includegraphics[width=1\textwidth]{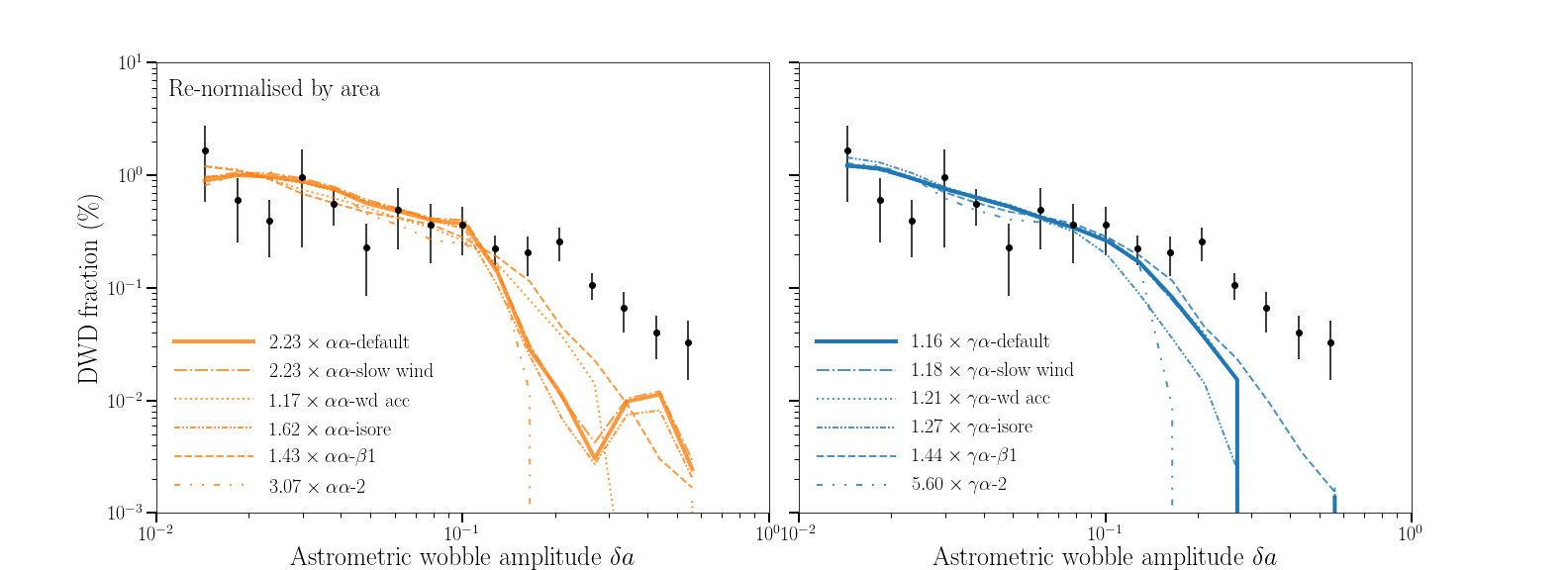}
\includegraphics[width=1\textwidth]{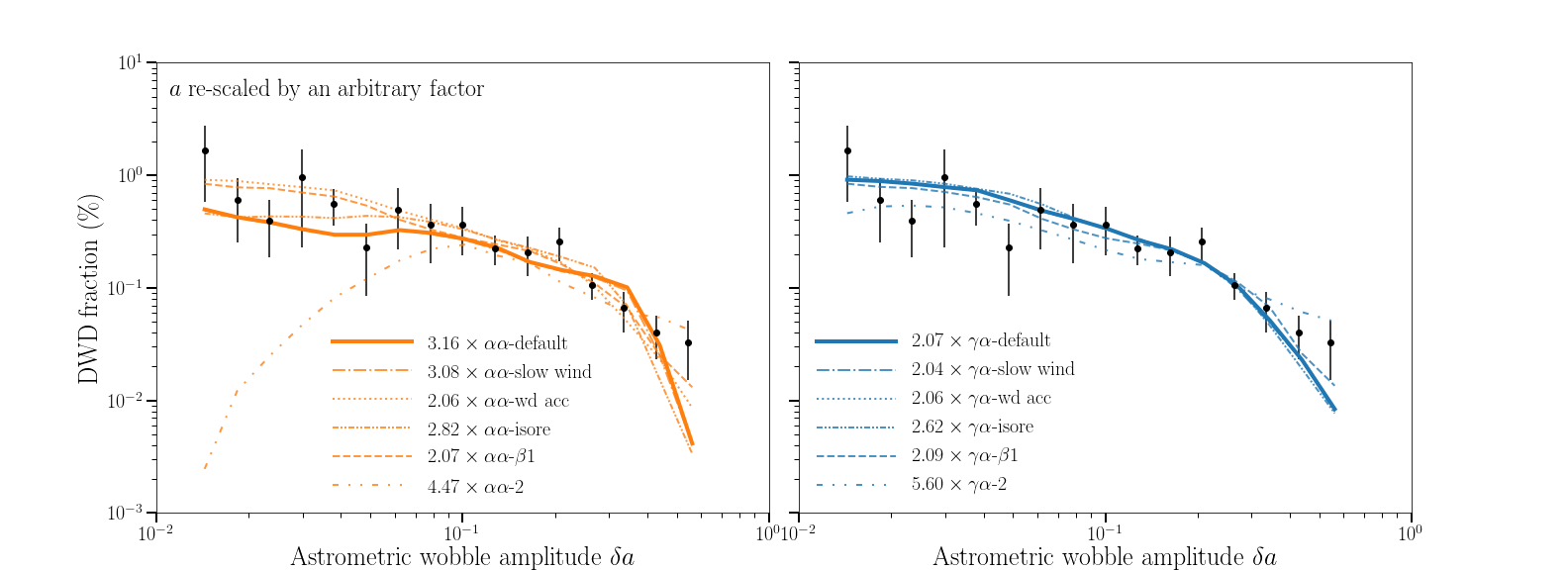}
  \caption{Theoretical distributions of the astrometric wobble amplitude as in the previous figure. Over-plotted as black circles the observed distribution of the astrometric wobble amplitude. Each model variation (each line) is multiplied by: an arbitrary multiplicative factor (top panels), a factor that accounts for a re-normalisation by area such that the total DWD fraction is equal to that estimated from the data (middle panels), a factor that re-scales binary separations to match the data (bottom panels).}
   \label{fig:bps_models_wobble_fit}
\end{figure*}

We consider the DWD fraction distribution with astrometric wobble amplitude and we fit 1) an arbitrary normalisation factor, 2) a normalisation factor such that the area under the model is equal (within error bars) to the measured DWD fraction, and 3) an overall factor that re-scales binaries' $\delta a$. Figure~\ref{fig:bps_models_wobble_fit} illustrates the results. As expected, changing the normalisation of the models makes up only for the total binary fraction (top and middle panels in Figure~\ref{fig:bps_models_wobble_fit}). However, a re-scaling factor for binaries' orbital separation can reconcile (most) models with the data (bottom panels Figure~\ref{fig:bps_models_wobble_fit}). Physically, this suggests that on average binaries' -- at least within the considered interval of orbital separations -- should shrink less. The obtained re-scaling factors for each model are reported in the legend of the Figure. We find that for the $\gamma \alpha$-family this factor is $\sim 2$, while for the $\alpha \alpha$-family it is $\sim 3$; this is excluding $\alpha\alpha$-2 and $\gamma\alpha$-2 models in which by construction the orbits shrink more than for the rest of the models.

\end{document}